\documentclass[twoside,11pt]{article}

\usepackage{jmlr2e_OW}
\usepackage{amsmath}
\usepackage[shortlabels]{enumitem}
\usepackage{bm}
\usepackage{booktabs}

\newtheorem{approximation}{Approximation} 
\newtheorem{assumption}{Assumption}

\makeatletter
\@addtoreset{approximation}{section}
\@addtoreset{approximation}{section*}
\makeatother

\newcommand{\appropto}{\mathrel{\vcenter{
  \offinterlineskip\halign{\hfil$##$\cr
    \propto\cr\noalign{\kern2pt}\sim\cr\noalign{\kern-2pt}}}}}

\jmlrheading{1}{2019}{1-48}{5/2018}{10/00}{meila00a}{Omer Weissbrod, Shachar Kaufman, David Golan and Saharon Rosset}

\ShortHeadings{Likelihood for GPs under Case-Control Sampling
}{Weissbrod, Kaufman, Golan and Rosset}
\firstpageno{1}

\begin{document}

\title{Maximum Likelihood for Gaussian Process Classification and Generalized Linear Mixed Models under Case-Control Sampling}

\author{
		\name Omer Weissbrod \email oweissbrod@hsph.harvard.edu \\
       \addr Department of Statistics and Operations Research\\
       School of Mathematical Sciences\\
       Tel-Aviv University\\
		  Tel-Aviv, Israel		  
	     \AND
       \name Shachar Kaufman \email shachark@post.tau.ac.il \\
       \addr Department of Statistics and Operations Research\\
       School of Mathematical Sciences\\
       Tel-Aviv University\\
		  Tel-Aviv, Israel
       \AND
       \name David Golan \email david@viz.ai \\
       \addr Viz.ai\\
		  Tel-Aviv, Israel    
       \AND
       \name Saharon Rosset \email saharon@post.tau.ac.il \\
       \addr Department of Statistics and Operations Research\\
       School of Mathematical Sciences\\
       Tel-Aviv University\\
		  Tel-Aviv, Israel
		  }

\editor{Ryan Adams}

\maketitle

\vspace{10pt}

\begin{abstract}
Modern data sets in various domains often include units that were sampled non-randomly from the population and have a  latent correlation structure. Here we investigate a common form of this setting, where every unit is associated with a latent variable, all latent variables are correlated, and the probability of sampling a unit depends on its response. Such settings often arise in case-control studies, where the sampled units are correlated due to spatial proximity, family relations, or other sources of relatedness. Maximum likelihood estimation in such settings is challenging from both a computational and statistical perspective, necessitating approximations that take the sampling scheme into account. We propose a family of approximate likelihood approaches which combine composite likelihood and expectation propagation. We demonstrate the efficacy of our solutions via extensive simulations. We utilize them to investigate the genetic architecture of several complex disorders collected in case-control genetic association studies, where hundreds of thousands of genetic variants are measured for every individual, and the underlying disease liabilities of individuals are correlated due to genetic similarity. Our work is the first to provide a tractable likelihood-based solution for case-control data with complex dependency structures.
\end{abstract}


\section{Introduction}

In the analysis of scientific data, a common phenomenon is the existence of complex dependencies between
analyzed units. This is encountered in diverse fields such as
epidemiology, econometrics, ecology, geostatistics, psychometrics and
genetics, and can arise due to spatial correlations, temporal
correlations, family relations, or other sources of heterogeneity
\citep{pfeiffer_spatial_2008,rabe-hesketh_maximum_2005,bolker_generalized_2009,rabe-hesketh_generalized_2004,yang_advantages_2014,burton_genetic_1999,diggle_model-based_1998}. This idea is often captured through the use of Gaussian processes (GPs; \citealt{rasmussen_gaussian_2006}) or equivalently, through generalized linear mixed models (GLMMs; \citealt{mcculloch_generalized_2008})
or latent Gaussian models \citep{fahrmeir_multivariate_2001}. Such models associate sampled units
with latent variables, and express the dependencies through covariance matrices of latent variables.

A second important concept is that of \emph{ascertainment}, where the probability of sampling a unit depends on its response. Ascertainment is especially common in
case-control studies where a binary response variable has a rare outcome, such as a rare disease, leading to oversampling of disease cases relative to their population prevalence \citep{breslow_statistics_1996}.

\begin{figure}
\centering
\includegraphics[width=0.75\linewidth]
{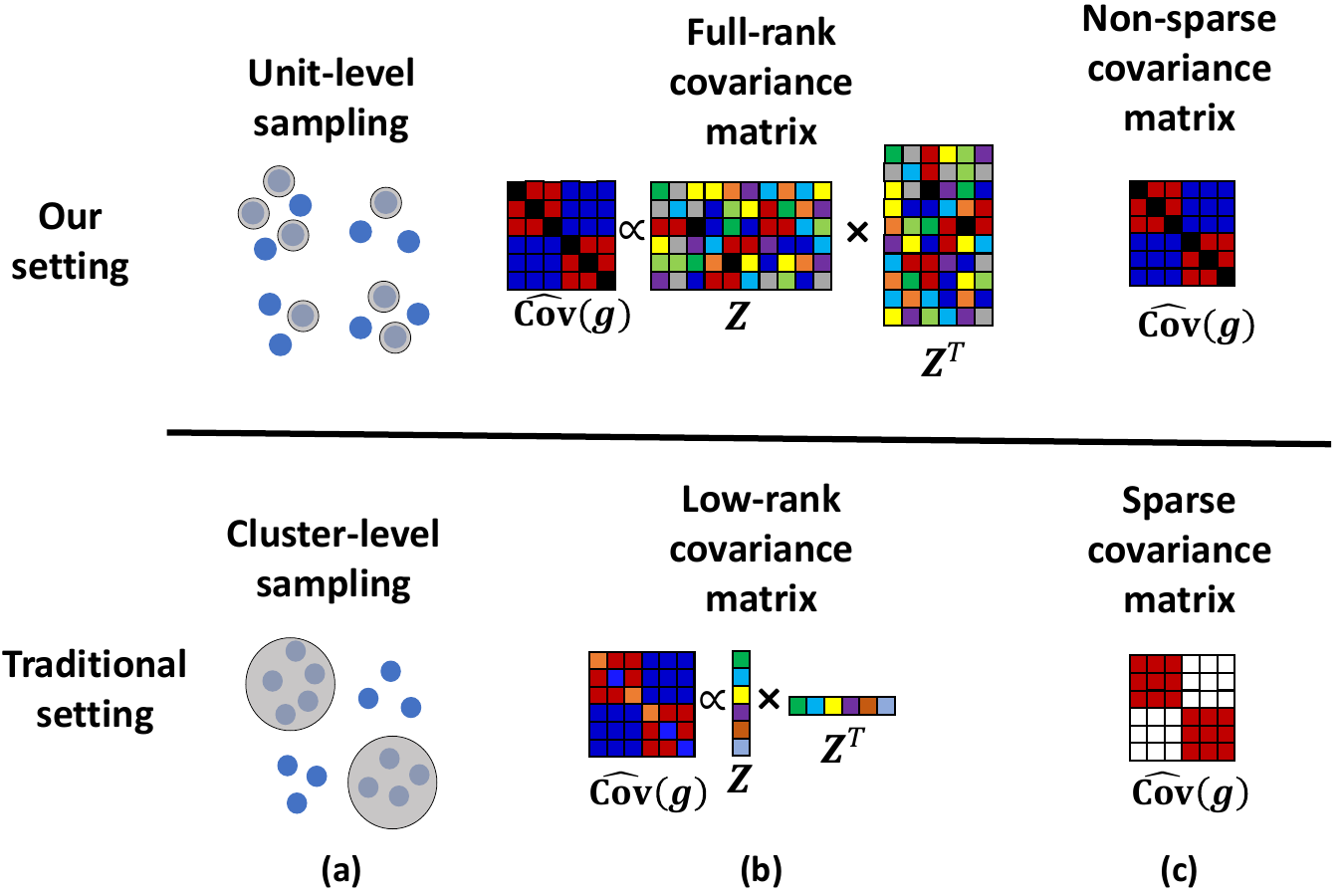}
\caption{\label{fig:classical_versus_us}
The properties that are unique to our setting of interest (top row) compared to more traditional settings (bottom row). (a) Unit-level sampling, where the decision whether to sample a unit depends only on their response, in contrast to studies that either sample or omit an entire cluster of units. (b) A full-rank covariance matrix of latent variables, indicating a complex dependency structure. (c) A non-sparse covariance matrix, indicating that the latent variables of every pair of units are correlated.}
\end{figure}

In this paper we consider situations that contain both elements---a
complex covariance structure and case-control sampling---and the
statistical modeling solutions available for these situations. Our
interest lies in an extreme form of this combination, involving:
\begin{enumerate}[(a)]
\item 
Unit-level ascertainment, where the sampling probability of a unit depends only on its response (Figure \ref{fig:classical_versus_us}a). This stands in contrast to common study designs such as family studies \citep{neuhaus_analysis_2002}, ascertained longitudinal studies \citep{liang_longitudinal_1986} or clustered case-control studies \citep{neuhaus_effect_1990}. In such studies each cluster is either entirely selected or entirely omitted from the study, such that the dependency structure in the sample and in the population are the same.

\item
A full-rank covariance matrix, indicating that dependencies cannot be captured by a small number of variables (Figure \ref{fig:classical_versus_us}b). Such settings are common in modern data sets due to either high-dimensional settings or to the use of kernels or basis expansions, which implicitly project a small number of features into a large (possibly infinite-dimensional) space \citep{diggle_model-based_1998,rasmussen_gaussian_2006}.

\item
A non-sparse covariance matrix, indicating that all units are correlated (Figure \ref{fig:classical_versus_us}c). Such settings exacerbate both the computational challenge, because the density does not factorize into multiplicative terms, and the statistical challenge, because classic statistical theory requires a large number of independent units.
\end{enumerate}
Special cases of this
combination have been addressed in the literature \citep{glidden_ascertainment_2002,epstein_ascertainment-adjusted_2002,neuhaus_family-specific_2006,neuhaus_likelihood-based_2014}, but to our
knowledge, there is a limited set of available solutions for the general
setting, which is indeed very challenging.

A major motivating application for our study is genome-wide association
studies of diseases with a case-control sampling design (GWAS-CC) \citep{price_progress_2015,visscher_10_2017}.
In GWAS-CC, the genomes of individuals affected with a disease and of
unaffected controls are collected in an effort to uncover the genetic
mechanisms driving disease risk.
Studies in this field have diverse goals,
reflected in the diversity of the statistical inference tasks they seek
to solve: testing for association between genetic variants and disease \citep{yang_advantages_2014, 	weissbrod_accurate_2015}, estimating disease heritability \citep{yang_common_2010, golan_measuring_2014, weissbrod2018estimating},
risk prediction \citep{zhou_polygenic_2013, golan_effective_2014,
weissbrod_multikernel_2016}, and more.

GWAS-CC typically employ case-control designs, where patients are
recruited in hospitals or clinics whereas healthy controls are recruited independently, owing to the small prevalence of complex genetic diseases---even common diseases like  type 1 diabetes or schizophrenia typically have a population prevalence $\leq$1\%. 
Furthermore, statistical models for such studies typically treat the effects of genetic variants on disease as random effects sampled from a distribution. This is because such studies include millions of variants, and the effects  are typically very small \citep{yang_common_2010,golan_measuring_2014, golan_effective_2014}. Hence, GWAS-CC give rise to settings with case-control sampling and a full-rank and non-sparse covariance structure: every individual has a latent genetic effect (given by the inner product of her genotype and the random effects), and the genetic effects are correlated due to genetic similarity.
Despite the extensive interest that GWAS-CC have attracted in recent years
\citep{wtccc_genome-wide_2007, ehret2011genetic,
sawcer_genetic_2011,
timmann2012genome,
ripke2014biological,
okada_genetics_2014},
the statistical modeling problems
this setting generates have been discussed in a limited manner, with
application of heuristic methods that do not formally take 
the probabilistic structure of the problem into account
\citep{lee_estimating_2011,hayeck_mixed_2015, weissbrod_accurate_2015, chen_control_2016,jiang_retrospective_2016}.

Similar settings arise in other scientific domains, where case-control
sampling and a full-rank, non-sparse covariance structure
are observed. Prominent examples include 
disease mapping studies with a smoothing kernel \citep{diggle_model-based_1998,kelsall_spatial_1998,held_towards_2005} and GP-based classification
of data collected in case-control studies \citep{chu_classification_2010,ziegler_individualized_2014,young_accurate_2013}. The analyses employed in these examples often
ignore the effects of case-control sampling, a practice we would like to
avoid and whose fundamental flaws we discuss and illustrate below.

The problem we consider poses substantial statistical and computational challenges. 
The main statistical framework for inference with binary responses and latent variables are GP classifiers, which are mathematically equivalent to latent Gaussian models, and recover GLMMs as a special case when using a linear kernel. Such models provide a likelihood-based solution but can pose significant computational difficulties.
Modern approach to alleviate computational difficulties include
(1) Pairwise likelihood (PL; \citealt{renard_pairwise_2004}), which approximates the joint distribution of all variables as a product of marginal distributions of pairs of variables;
(2) Expectation propagation (EP) \citep{minka_expectation_2001, seeger2005expectation}, which replaces multiplicative terms in the distribution with simpler terms from an exponential family distribution;
(3) Variational approximation \citep{opper_variational_2009,hensman2015mcmc}, which approximates the distribution with the closest distribution from a more tractable class;
(4) Markov chain Monte Carlo (MCMC) sampling combined with thermodynamic integration
\citep{kuss_assessing_2005,nickisch_approximations_2008,gelman_simulating_1998}, and
(5) Laplace approximation \citep{tierney_accurate_1986, raudenbush_maximum_2000} and its close variant, penalized quasi likelihood approximation \citep{breslow_approximate_1993,wolfinger_generalized_1993}, which approximate the distribution as a Gaussian distribution via a second-order Taylor expansion.
Of these, PL and EP have proved to often outperform  the alternatives  \citep{kuss_assessing_2005,nickisch_approximations_2008,varin_overview_2011} and thus form the basis of our proposed approach.

The main approach for  statistical inference in the presence of ascertainment is called \textit{ascertained maximum likelihood} (AML). AML consists of defining a binary variable indicating inclusion in the study for every unit, and conditioning the analysis on these variables \citep{scott_maximum_2001}. However, none of the above GP approximation methods is suitable for AML inference. The only existing method that is both scalable and statistically sound 
is a method-of-moments approach called phenotype-correlation-genotype-correlation (PCGC; \citealt{golan_measuring_2014}). However, this approach is not likelihood-based and thus cannot naturally be used for model comparison, inference, prediction, and hypothesis testing.

Here we propose two approaches for approximate likelihood computation in our setting of interest. Our approaches combine (1) GP + AML + PL, which provides a tractable likelihood approximation
but is very sensitive to model misspecification; and (2) GP + AML + modified EP, which 
is more computationally intensive, but is more robust and is closer to  traditional maximum likelihood estimation. We evaluate the merits of our approaches on both synthetic and real data sets of genetic studies involving thousands of individuals and hundreds of thousands of features treated as having random effects.

\section{Formal Problem Description}
Here we provide a formal description of our problem and the statistical and computational challenges.

\subsection{Gaussian Processes / Generalized Linear Mixed Models}
We are presented with $n$ units, with each unit $i$ having $c$ features $\bm{X}_{i} \in \mathbb{R}^c$ associated with fixed (non-random) effects
$\bm{\beta} \in \mathbb{R}^c$, 
$m$ additional features $\bm{Z}_{i} \in \mathbb{R}^{m}$ that are not associated with fixed effects,
and an outcome variable
$y_{i}$ (here we consider binary outcomes, but the framework can also be applied to other types).
We associate every unit $i$ with a latent variable $g_i(\bm{Z}_i)$ that depends on $\bm{Z}_{i}$ such that
$
P \left( y_{i}=1\ |\ \bm{X}_{i}, g_i, \bm{\beta} \right) =  h\left( \bm{X}_{i}^{T}\bm{\beta} + g_i \right), 
$
where $h\left( \cdot \right)$ is a likelihood function (which is closely related to an inverse link function in GLMM terminology), such as probit or logit. 

GPs impose a Gaussian process prior over the the functional form of the latent variables,
$
g \sim \mathcal{GP}\left(\bm{0}, k(\cdot, \cdot \,;\, \bm{\theta}) \right), \nonumber
$
where $k(\cdot, \cdot \,;\, \bm{\theta})$ is a \textit{kernel function} parameterized by $\bm{\theta}$. This indicates that
for every finite set of units $i=1 \ldots n$, the vector $\bm{g}=\left\lbrack g_{1}(\bm{Z}_1), \ldots, g_{n}(\bm{Z}_n), \right\rbrack^T$ follows a multivariate normal distribution:
\begin{align}
\bm{g} \sim \mathcal{N}\left(\bm{0}, \bm{K}(\bm{Z}; \bm{\theta})\right), \nonumber
\end{align}
where $\bm{K}$, which is non-sparse and full-rank in our setting, has entries $K_{ij} = k(\bm{Z}_i, \bm{Z}_j; \bm{\theta})$.

\begin{figure}
\centering
\includegraphics[width=0.65\linewidth]
{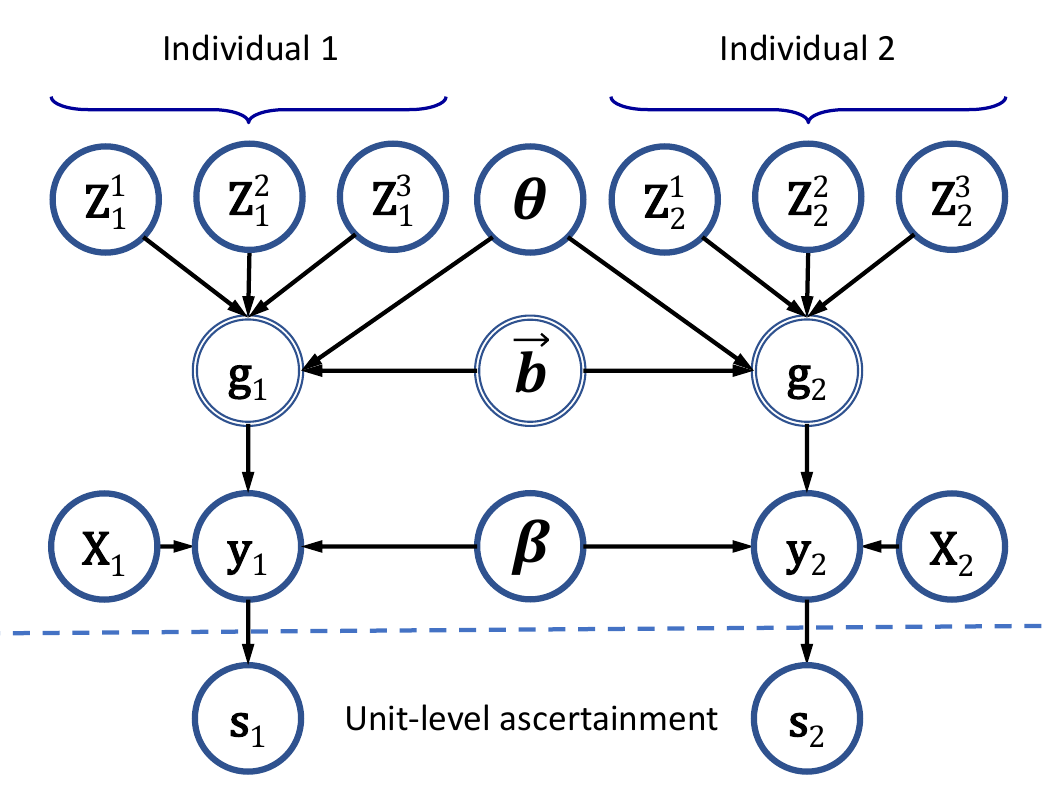}
\caption{\label{fig:graphical_model}
A directed graphical model for a GP with one feature $X$ associated with a fixed effect $\beta$, three features $Z^1$, $Z^2$, $Z^3$ associated with an implicit vector of random effects $\vec{\bm{b}}$ and with a hyperparameter $\theta$, and two sampled units (indicated by  subscript indices) with latent variables $g_1$, $g_2$ and observed responses $y_1$, $y_2$. Also shown is the extension to unit-level ascertainment, which consists of adding a sampling indicator $s_i$ that depends on $y_i$ and is equal to 1 for every sampled unit. Latent (non-observed) random variables are marked with a double-lined border.
}
\end{figure}

GLMMs are a class of models that are mathematically equivalent to GPs with linear kernels, i.e., $\bm{K}_{ij} = \theta \bm{Z}_i^T \bm{Z}_j$, where $\theta$ is a scalar hyperparameter called a \emph{variance component}. 
GLMM literature typically uses the alternative notation \(g_{i} = \bm{Z}_{i}^{T}\bm{b}\), where 
\(\bm{b} \sim \mathcal{N}\left(\bm{0}, \sqrt{\theta} \bm{I} \right)\) is a vector of \textit{random effects}. The equivalence can be extended to non-linear kernels by projecting $\bm{Z}_i$ into a high (possibly infinite) dimensional space that depends on $\bm{\theta}$, and defining \(\bm{b} \sim \mathcal{N}\left(\bm{0}, \bm{I} \right)\) \citep{rasmussen_gaussian_2006}.
GLMMs can also be defined with non-normal random effects, in which case the equivalence with GPs breaks down, but we do not consider such models here.
We focus on linear kernels because we are interested in extremely high-dimensional settings, where non-linear kernels often overfit \citep{weissbrod_multikernel_2016}. 
A schematic graphical model of GPs and GLMMs is shown in Figure \ref{fig:graphical_model}.

Our main aim in this work is estimating the kernel hyperparameters $\bm{\theta}$ (i.e., training the model). Given a vector of observed outcomes
\(\bm{y} = \left\lbrack y_1, \ldots, y_n \right\rbrack^T\) and the matrices
\(\bm{X} = \left\lbrack \bm{X}_{1},\ldots,\bm{X}_{n} \right\rbrack^{T}\),
\(\bm{Z} = \left\lbrack \bm{Z}_{1},\ldots,\bm{Z}_{n} \right\rbrack^{T}\),
the GP likelihood is given by:
\begin{align}
L\left( \bm{\beta},\bm{\theta} \right) = P\left( \bm{y}\ |\ \bm{X},\bm{Z}, \bm{\theta}, \bm{\beta} \right) & = 
\int{P\left( \bm{g|Z},\bm{\theta} \right)
\prod_{i}
{P(y_{i}|\bm{X}_{i},g_{i},\bm{\beta})} 
d\bm{g}}. \label{eq:GP_likelihood}
\end{align}
As we do not impose a prior over the hyperparameters $\bm{\beta},\bm{\theta}$, we can estimate them via type-II maximum likelihood, by finding the values that maximize Equation \ref{eq:GP_likelihood}.

\subsection{The Implications of Ignoring Ascertainment in GPs}

Up until now we implicitly assumed that $n$ units were sampled completely at random from an underlying population. We now assume that the sample is ascertained, i.e., that the probability of sampling cases (units with $y_i=1$) and controls ($y_i=0$) is different (see Figure \ref{fig:graphical_model}). 

We first demonstrate that using GPs while ignoring ascertainment leads to nonsensical conclusions which stand in contrast to  fundamental motivations for GP use, like the central limit theorem. We focus on binary GPs, which can be formulated according to the
liability threshold model \citep{dempster_heritability_1950}. Under this model, every unit $i$ has a
latent liability \(l_{i} = g_{i} + \epsilon_{i}\), where
\(\epsilon_{i}\) is an iid latent residual variable whose distribution
depends on the likelihood function (e.g. normally distributed for probit,
or logit distributed for logit), and unit $i$ is a case (having $y_i=1$) if and only if
\(l_{i} > t\) for some cutoff \(t\). The \textit{prevalence} $K$ is the proportion of units in the population having $l_i>t$. 
We emphasize that a normally distributed $\epsilon_i$ is completely equivalent to a standard GP with a probit likelihood $h(\cdot)$. 

\begin{figure}
\centering
\includegraphics[width=1\linewidth]
{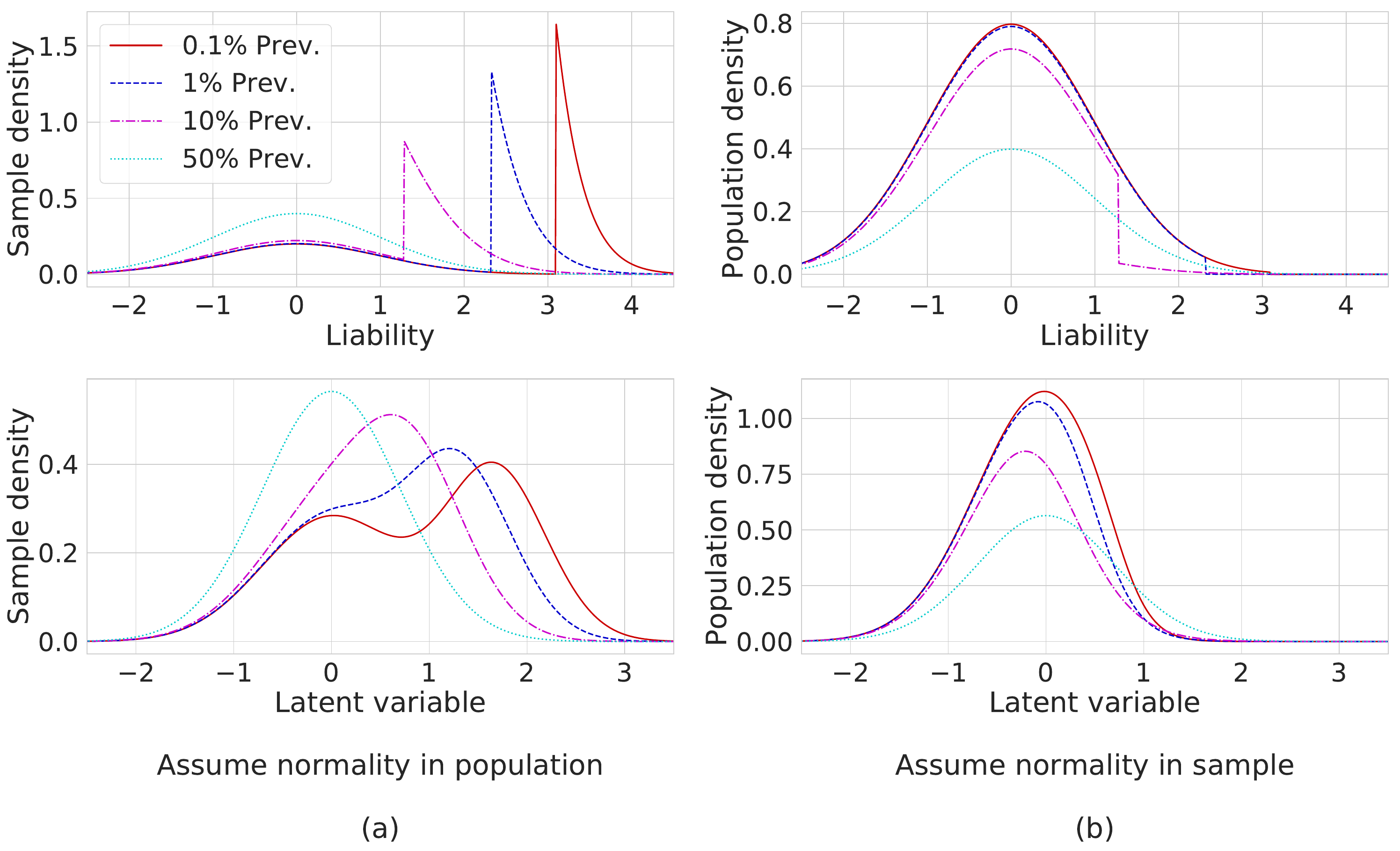}
\caption{\label{fig:fig_densities}
\vspace{-12pt}
The implications of assuming normality of latent variables in the population from which units are sampled (panel a) or in a case-control study (panel b), for a GP with a probit likelihood and a sample consisting of 50\% cases. The liability is given by $l_i=g_i+\epsilon_i$, $g_i, \epsilon_i \sim \mathcal{N}(0,\sqrt{0.5})$. Units with liabilities greater than their (1-prevalence) population quantile are cases. (a) When assuming normality in the population, latent variables and liabilities in a case-control study are not normally distributed (unless the cases prevalence is 50\%, in which case there is no ascertainment). (b) When naively assuming normality of latent variables in a case-control study, the latent variables and the liabilities are \emph{not} normally distributed in the population from which the data was sampled, in contradiction to the liability threshold model. Specifically, the liabilities distribution is discontinuous, and the latent variables distribution has a heavy left tail. All distributions were computed analytically by conditioning on the sampling indicators defined in Section \ref{sec:ascertainment}.}
\end{figure}

It is common to use likelihood functions associated with a smooth and
symmetrically distributed \(\epsilon_{i}\), such as logit or probit,
which leads to a smooth and symmetric distribution of liabilities in the
population. However, due to the ascertainment mechanism, the liabilities and  latent variables \(g_{i}\)
in an ascertained sample follow a non-symmetric and possibly
discontinuous distribution (Figure \ref{fig:fig_densities}a), and thus cannot be analyzed
with standard likelihood functions. This problem motivates our proposed solutions for analysis of case-control studies.
Many studies in practice ignore the complexities above, and
instead use common likelihood functions such as a logit or a probit in
case-control studies (e.g.  \citealt{chen_control_2016,jiang_retrospective_2015,kramer_african_2017,qi_genetics_2017}). However, this solution implies a
non-symmetric and discontinuous distribution of
latent variables in the population from which units are sampled, in
stark contrast to the central limit theorem assumptions (Figure \ref{fig:fig_densities}b). Thus,
ignoring the ascertainment scheme in GPs may lead to nonsensical
probabilistic settings under common assumptions.

The practical implications of modeling violations due to ascertainment have been investigated extensively in the statistical genetics literature. These include severe biases in estimation of quantities such as disease heritability \citep{golan_measuring_2014,weissbrod2018estimating}, inaccurate risk prediction \citep{golan_effective_2014} and loss of power in hypothesis testing \citep{weissbrod_accurate_2015,hayeck_mixed_2015,yang_advantages_2014}.

\subsection{Modeling Ascertainment in GPs \label{sec:ascertainment}}

Analysis of ascertained data is typically performed by (1) defining a binary sampling indicator $s_i$ for every unit $i$ such that $s_i$ depends only on $y_i$ (Figure \ref{fig:graphical_model}); and (2) performing all statistical inference tasks conditional on \(s_{1} = 1,\ldots,s_{n} = 1\). This requires specifying the sampling probabilities of cases $P(s_i=1|y_i=1)$ and of controls $P(s_i=1|y_i=0)$. There are two main approaches for estimating the hyperparameters $\bm{\theta}_{\bm{y}|\bm{X},\bm{Z}}$ of the distribution $P\left( \bm{y}|\bm{X},\bm{Z} \right)$ with such indicators, differing with respect to how the sampling probabilities are determined.

\textit{Maximum profile likelihood} estimates $\bm{\theta}_{\bm{y}|\bm{X},\bm{Z}}$ by jointly maximizing the so-called profile likelihood $P\left( \bm{y}|\bm{X},\bm{Z}, s_1=1, \ldots,s_n=1 \right)$ over both $\bm{\theta}_{\bm{y}|\bm{X},\bm{Z}}$ and the nuisance hyperparameters $\bm{\theta}_{s|y}$ of the distributions 
\(P\left( s_{i} = 1|y_{i} \right)\) of every possible value of
\(y_{i}\),
 i.e.,
 $\hat{\bm{\theta}}_{\bm{y}|\bm{X},\bm{Z}}$, $\hat{\bm{\theta}}_{s|y}$=
$\text{argmax}_{\bm{\theta}_{\bm{y}|\bm{X},\bm{Z}}, \bm{\theta}_{s|y}}$
$P\left(  \bm{y}\ |\ \bm{X},\bm{Z},s_1=1, \ldots,s_n=1, \bm{\theta}_{\bm{y}|\bm{X},\bm{Z}},\bm{\theta}_{s|y} \right)$
\citep{scott_maximum_2001}.
The resulting estimator is maximally efficient as it attains the Cram\'er-Rao lower bound.

\textit{Ascertained Maximum Likelihood} (AML) estimates $\bm{\theta}_{\bm{y}|\bm{X},\bm{Z}}$ using a pre-specified assignment $\bm{\theta}_{s|y}=\bm{\theta}_{s|y}^0$,
i.e., 
 $\hat{\bm{\theta}}_{\bm{y}|\bm{X},\bm{Z}}$ =
$\text{argmax}_{\bm{\theta}_{\bm{y}|\bm{X},\bm{Z}}}$
$P\left(  \bm{y}\ |\ \bm{X},\bm{Z},s_1=1, \ldots,s_n=1, \bm{\theta}_{\bm{y}|\bm{X},\bm{Z}},\bm{\theta}_{s|y}^0 \right)$
\citep{scott_fitting_1997}. For a
binary outcome with a population prevalence $K$ and an in-sample
prevalence $P$, any assignment $\bm{\theta}_{s|y}^0$ obeying the constraint
\(\frac{P\left( s_{i} = 1|y_{i} = 0 \right)}{P\left( s_{i} = 1|y_{i} = 1 \right)} = \frac{K\left( 1 - P \right)}{\left( 1 - K \right)P}\)
guarantees consistent estimates (i.e., estimators converge to the true parameter values as sample size tends to infinity if the model is true), because it yields the observed
case-control ratio in expectation. This approach is often termed pseudo
likelihood or conditional likelihood \citep{manski_alternative_1981,hsieh_estimation_1985}, but as both terms have
alternative meanings in GLMM literature, we use the term ascertained
likelihood instead.
AML is less statistically efficient than  maximum profile likelihood, in the sense that the estimator has a larger variance. However, the loss of efficiency has been shown to be negligible in practice \citep{wild_fitting_1991,scott_fitting_1997}. AML  has previously been used for family-based studies \citep{glidden_ascertainment_2002,epstein_ascertainment-adjusted_2002}, but to our knowledge it has not been used under the combination of a dependency
structure and unit-level sampling.

To combine GPs with the AML framework, we define the ascertained GP
likelihood and apply Bayes' law as follows:
\begin{align}
L^{*}\left( \bm{\beta},\bm{\theta} \right) = P\left( \bm{y}\ |\ \bm{X},\bm{Z},\bm{s} = 1, \bm{\theta},\bm{\beta} \right) = \frac{P\left( \bm{y}\ |\ \bm{X},\bm{Z}, \bm{\theta},\bm{\beta} \right)}{P\left( \bm{s = 1}\ |\ \bm{X},\bm{Z}, \bm{\theta},\bm{\beta} \right)}\prod_{i}^{}{P\left( s_i = 1 | y_i \right)}, \label{eq:combine_glmm_aml}
\end{align}
where 
$\bm{s}=1$ is a shorthand notation for $s_1=1, \ldots,s_n=1$ and
$\bm{\theta}_{\bm{y}|\bm{X},\bm{Z}} = \left\{\bm{\theta}, \bm{\beta}\right\}$.
The last term in the rhs of Equation \ref{eq:combine_glmm_aml} is considered known under AML
and requires no special treatment. The numerator is the likelihood of a 
standard GP under no ascertainment, and the
denominator is the likelihood of a GP in which the outcome is
\(s_{i}\) instead of \(y_{i}\). A naive approach is to approximate the
numerator and denominator separately. However, obtaining an accurate
estimate of the ratio is extremely challenging, because both the
numerator and denominator are challenging to approximate, and any
inaccuracy is compounded by the division. In our experience, this
approach does not lead to reasonable estimators.

\section{Approximate Inference in GPs under Ascertainment \label{sec:approx_inference}}

We now propose two methods for approximate inference in GPs under ascertainment.

\subsection{Ascertained Pairwise Likelihood \label{subsec:PL}}

PL is composite likelihood approximation, which
approximates a multivariate joint density via
a product of marginal densities of pairs of variables \citep{varin_overview_2011}:
\begin{align}
P\left( \bm{y}\ |\ \bm{X},\bm{Z},\bm{\theta},\bm{\beta} \right)&  \appropto
\prod_{i \neq j}
{P\left( y_{i},y_{j}\ |\ \bm{X}_{i},\bm{X}_{j},\bm{Z}_{i},\bm{Z}_{j}, \bm{\theta}, \bm{\beta} \right)}, \label{eq:PL}
\end{align}
where $i$,$j$ iterate over all pairs of units, and $\appropto$ indicates approximate proportionality with respect to the  hyperparameters $\bm{\theta}$, $\bm{\beta}$. The maximum pairwise likelihood estimate is approximately the maximum likelihood estimate.
PL is computationally
efficient owing to its quadratic dependency on the sample size, and is consistent under suitable regularity conditions \citep{varin_overview_2011}.

Ascertained PL (APL) is an extension of PL that approximates the ascertained likelihood in Equation \ref{eq:combine_glmm_aml} by modifying
Equation \ref{eq:PL} to condition on $\bm{s}=1$ :
\begin{align}
P\left( \bm{y}\ |\ \bm{X},\bm{Z,s = 1} \right) & \appropto
\prod_{i \neq j}
{P\left( y_{i},y_{j}\ |\ \bm{X}_{i},\bm{X}_{j},\bm{Z}_{i},\bm{Z}_{j},s_{i} = 1,s_{j} = 1 \right)}  \nonumber \\
& = \ \prod_{i \neq j}\frac
{
P\left( y_{i},y_{j}\ |\ \bm{X}_{i},\bm{X}_{j},\bm{Z}_{i},\bm{Z}_{j} \right)
}
{
P\left( s_{i} = s_{j} = 1\ |\ \bm{X}_{i},\bm{X}_{j},\bm{Z}_{i},\bm{Z}_{j} \right)
}
P\left( s_{i} = s_{j} = 1|\ y_{i},y_{j} \right), \nonumber
\end{align}
where
\(P\left( s_{i} = s_{j} = 1|\ y_{i},y_{j} \right) = P\left( s_{i} = 1|\ y_{i} \right)P\left( s_{j} = 1|\ y_{j} \right)\)
are known constants which can be ignored, and we omitted the hyperparameters $\bm{\beta}$, $\bm{\theta}$ for brevity. The terms in the numerator
and the denominator can be separately evaluated as in standard PL, where
we treat the denominator as a GP with a suitable likelihood function.
Unlike Equation \ref{eq:combine_glmm_aml}, the evaluation of the ratio is accurate since both the numerator
and denominator can be computed exactly. In certain settings, PL
evaluation can be substantially accelerated via a Taylor approximation
around \(\bm{Z}_{i}^{T}\bm{Z}_{j} = 0\), which enables factoring
each bivariate distribution into a product of marginal distributions
(Appendix A).

\subsection{Ascertained Expectation Propagation \label{subsec:EP}}
EP is a popular approach for approximating complex distributions by iteratively replacing every multiplicative term in the joint distribution of the observed and latent variables with a simpler term from an exponential family distribution \citep{minka_expectation_2001,rasmussen_gaussian_2006,seeger2005expectation}.  This joint distribution in GPs is given by
\(P\left( \bm{g|Z},\bm{\theta} \right)\prod_{i}{P(y_{i}|\bm{X}_{i},g_{i},\bm{\beta})}\).
EP replaces every term in this product by an unnormalized Gaussian, 
$
P\left( y_{i}|{\bm{X}_{i},g}_{i} \right) \approx t_{i}\left( g_{i} \right) \triangleq r_{i}\mathcal{N}\left( g_{i};{\widetilde{\alpha}}_{i},{\widetilde{\gamma}}_{i} \right)
$,
where we omitted the hyperparameters $\bm{\beta}$,  $\bm{\theta}$ for brevity, and the site parameters $r_i$, $\widetilde{\alpha}_i$, $\widetilde{\gamma}_i$ implicitly depend on $\bm{X}_i$, $y_i$ and $\bm{\beta}$. EP iteratively updates the terms $t_i(g_i)$, such that each term minimizes the generalized Kullback Leibler divergence (GKL) between the functions $q_{-i}(g_i)t_i(g_i)$ and $q_{-i}(g_i)P(y_i | \bm{X}_i, g_i)$  (i.e., the KL divergence between these functions after standardizing them to integrate to unity), where
the cavity distribution $q_{-i}(g_i) \propto \int{P(\bm{g}|\bm{Z}) \prod_{j \neq i}t_j(g_j) d\bm{g}_{j \neq i}}$
represents the current approximation of $P\left(g_i | \bm{Z}, \bm{y}_{j \neq i}\right)$.

Given an EP approximation, the GP likelihood can be approximated as:
\begin{align}
P\left( \bm{y}\ |\ \bm{X},\bm{Z} \right) & \approx 
\int{P\left( \bm{g|Z} \right)\prod_{i}{t_{i}\left( g_{i} \right)}d\bm{g}}. \nonumber
\end{align}
This expression can be evaluated analytically because it is an integral
of a product of (unnormalized) Gaussian densities. EP has proven to consistently outperform alternative 
approximation methods for binary data \citep{nickisch_approximations_2008}, and recent theoretical
analysis has demonstrated that is is statistically consistent under certain modeling
assumptions \citep{dehaene_bounding_2016,dehaene2018expectation}.

Ascertained EP (AEP) is our proposed method to generalize standard EP to handle ascertainment. AEP 
approximates the ascertained likelihood in Equation \ref{eq:combine_glmm_aml} by replacing the standard EP  step with a modified step that equates the functions $\int q_{-i}(g_i)t_i(g_i) dg_i$ and
$
\frac{\int q_{-i}(g_i)P(y_i,s_i|g_i,\bm{X}_i)dg_i}{\int q_{-i}(g_i)P(s_i|g_i,\bm{X}_i)dg_i}
$.
Unlike standard EP, we cannot minimize the GKL divergence beteween these functions, because this will lead to the same solution as standard EP, up to a scaling constant. Instead, AEP finds the unnormalized Gaussian $t_i(g_i)$ which makes these functions and their first two partial derivatives with respect to $\mu_{-i}$ (the mean of the Gaussian $q_{-i}(g_i)$) have the same value when evaluated at $\mu_{-i}$ (see Appendix B for details). 

EP is a special case of AEP, because the proposed step objective coincides with the standard EP objective in the absence of ascertainment (i.e., when $P(s_i | y_i)$ is a constant regardless of $y_i$). To see this, observe that EP minimizes the GKL divergence between  
$\hat{m}(g_i) \triangleq q_{-i}(g_i)  P(y_i | g_i,\bm{X}_i)$
and
$
\tilde{m}(g_i) \triangleq q_{-i}(g_i) \cdot t_i(g_i)
$.
Since $\tilde{m}(g_i)$ is an unnormalized Gaussian, EP minimizes the GKL divergence by equating its zeroth, first and second moments with those of $\hat{m}(g_i)$ \citep{rasmussen_gaussian_2006}. Hence, standard EP requires computing the mean $\hat{\mu}_i$ and variance $\hat{\sigma}^2_i$ of $\hat{m}(g_i)$.
A straightforward but lengthy derivation shows that we can compute these quantities as follows:
\begin{align}
\hat{\mu}_i& = \frac{\partial}{\partial \mu_{-i}}\left[\log \int{\hat{m}(g_i) dg_i}\right]\sigma^2_{-i} + \mu_{-i}  \nonumber \\
\hat{\sigma}^2_i & = \frac{\partial ^2}{\partial \left(\mu_{-i}\right)^2}\left[\log \int{\hat{m}(g_i) dg_i}\right]\left(\sigma_{-i}^2\right)^2 + \sigma^2_{-i},  \nonumber  
\end{align}
where $\mu_{-i}$, $\sigma^2_{-i}$ are the mean and variance of the Gaussian $q_{-i}(g_i)$, and the derivatives are evaluated at the actual value of $\mu_{-i}$. Hence, there is a one-to-one correspondence between the first two moments of $\hat{m}(g_i)$ and its first two partial derivatives with respect to $\mu_{-i}$ (when evaluated at $\mu_{-i}$). Consequently, each step of standard EP can alternatively be described as imposing the constraint that the zeroth, first and second derivatives of the integrals of $\tilde{m}(g_i)$ and $\hat{m}(g_i)$ with respect to $\mu_{-i}$ are the same. This is the same constraint used in AEP. Hence, EP and AEP coincide in the absence of ascertainment, where $P(s_i=1|y_i)$ is constant regardless of the value of $y_i$.

The sampling variance of the AEP maximum likelihood estimator can be estimated efficiently via jackknife sampling, by  reusing the functions $t_i(g_i)$ \citep{opper_gaussian_2000,qi_predictive_2004,vehtari_bayesian_2016}.
The evaluation of each jackknife sample requires inverting a matrix that is a submatrix of a matrix that was inverted in the original computation, with one row and one column removed. Such an inversion can be computed rapidly while retaining numerical stability, by combining a Cholesky decomposition with a series of Givens rotations \citep{seeger2004low}.
A formal analysis of AEP is difficult because there are relatively few theoretical guarantees for standard EP, which is a special case of AEP, except under relatively strong assumptions \citep{dehaene_bounding_2016, dehaene2018expectation}. However, we sketch a heuristic argument supporting the objective function of AEP in Appendix B.

\section{Results}

We evaluated the performance of our methods using extensive simulations and real data analysis. We first describe our simulation studies, and then present the results obtained on real data.

\subsection{Simulations Overview}
We simulated data that closely mimics real GWAS-CC using the liability threshold model, where each individual has liability
\(l_{i} = \bm{X}_{i}^{T}\bm{\beta} + g_i + \epsilon_{i}\), $g_i$ is a GP latent variable,
and individuals with $l_i$ greater than some cutoff are cases.
In most simulations we used a linear kernel with a single hyperparameter, $\bm{\theta}=\{\sigma^2_g\}$, though we also investigate radial basis function (RBF) kernels below. 
Our aim was estimating the hyperparameter $\sigma^2_g$, which we call a variance component.
Importantly, 
$\sigma^2_g / \text{var}(l_i)$ 
is an estimator of genetic heritability, defined as the proportion of $\text{var}(l_i)$ explained by genetics.

We simulated genetic data based on single nucleotide polymorphisms (SNPs), which can be encoded as 0/1/2, according to the number of minor alleles carried by an individual at a specific position in the genome. We first generated a minor allele frequency $f^j \sim \mathcal{U}(0.05,0.5)$ for every SNP $j$, and then sampled a matrix $\bm{Z}$ of SNP, such that $Z_{ij} \sim \text{Bin}(2,f^j)$. Finally, we standardized each column in the matrix $\bm{Z}$ by subtracting the mean and dividing by the standard deviation corresponding to its allele frequency.

To simulate unit-level ascertainment, we (1) generated a population of
1,000,000 individuals, where for every individual $i$ we generated a vector of standardized genotypes $\bm{Z}_{i} \in \mathbb{R}^m$ as described above, and a vector 
 \(\bm{X}_{i}\mathcal{\sim \mathcal{N}}\left( 0,\bm{I} \right) \in \mathbb{R}^{c}\)
representing additional standardized risk factors such as sex or age;
(2) generated vectors
\(\bm{b}\mathcal{\sim \mathcal{N}}\left( 0,\sqrt{\sigma_{g}^{2}/m}\bm{I} \right)\)
of random effects, and
\(\bm{\beta}\mathcal{\sim \mathcal{N}}\left( 0,\sqrt{\sigma_{c}^{2}/c}\bm{I} \right)\)
of fixed effects; (3) assigned a latent variable
\(g_{i} = \bm{Z}_{i}^{T}\bm{b}\) and a liability
\(l_{i} = g_{i} + \bm{X}_{i}^{T}\bm{\beta}{+ \epsilon}_{i}\) to
every individual \(i\), where
\(\epsilon_{i}\mathcal{\sim \mathcal{N}}\left( 0,\sqrt{1 - \sigma_{c}^{2} - \sigma_{g}^{2}} \right)\)
iid; (4) defined all individuals with \(l_{i}\) greater than the
\(1 - K\) quantile of the liability distribution as cases (i.e., $y_i=1$), where \(K\)
is the desired prevalence; and (5) selected a subset of \(\frac{n}{2}\)
cases and \(\frac{n}{2}\) controls for the case-control study, where
\(n\) is the desired study size. Unless stated otherwise, we used
\(m = 500\), \(n = 500\), \(K = 1\%,\)
\(\sigma_{g}^{2} = 0.25, \sigma_{c}^{2} = 0.25\), and $c=1$. We generated 100 datasets for each unique combination of evaluated settings.

We evaluated three likelihood-based methods: AEP, APL and plain EP, which does not account for ascertainment. We additionally evaluated the moment-based method PCGC, which is considered the state of the art approach for hyperparameter estimation in genetic case-control studies \citep{golan_measuring_2014}. We estimated hyperparameters in the likelihood-based methods via maximum likelihood, and in PCGC by finding the values that minimize the squared loss between the observed and expected values of $\textrm{cov}(y_i, y_j)$ across all pairs of individuals $i,j$.
In all settings, we first estimated the fixed effects via a novel ascertainment-aware generalized estimation equations (AGEE) approach that we developed, and then adjusted the affection cutoffs accordingly (Appendix C). 

\begin{figure}
\centering
\includegraphics[width=1\linewidth]
{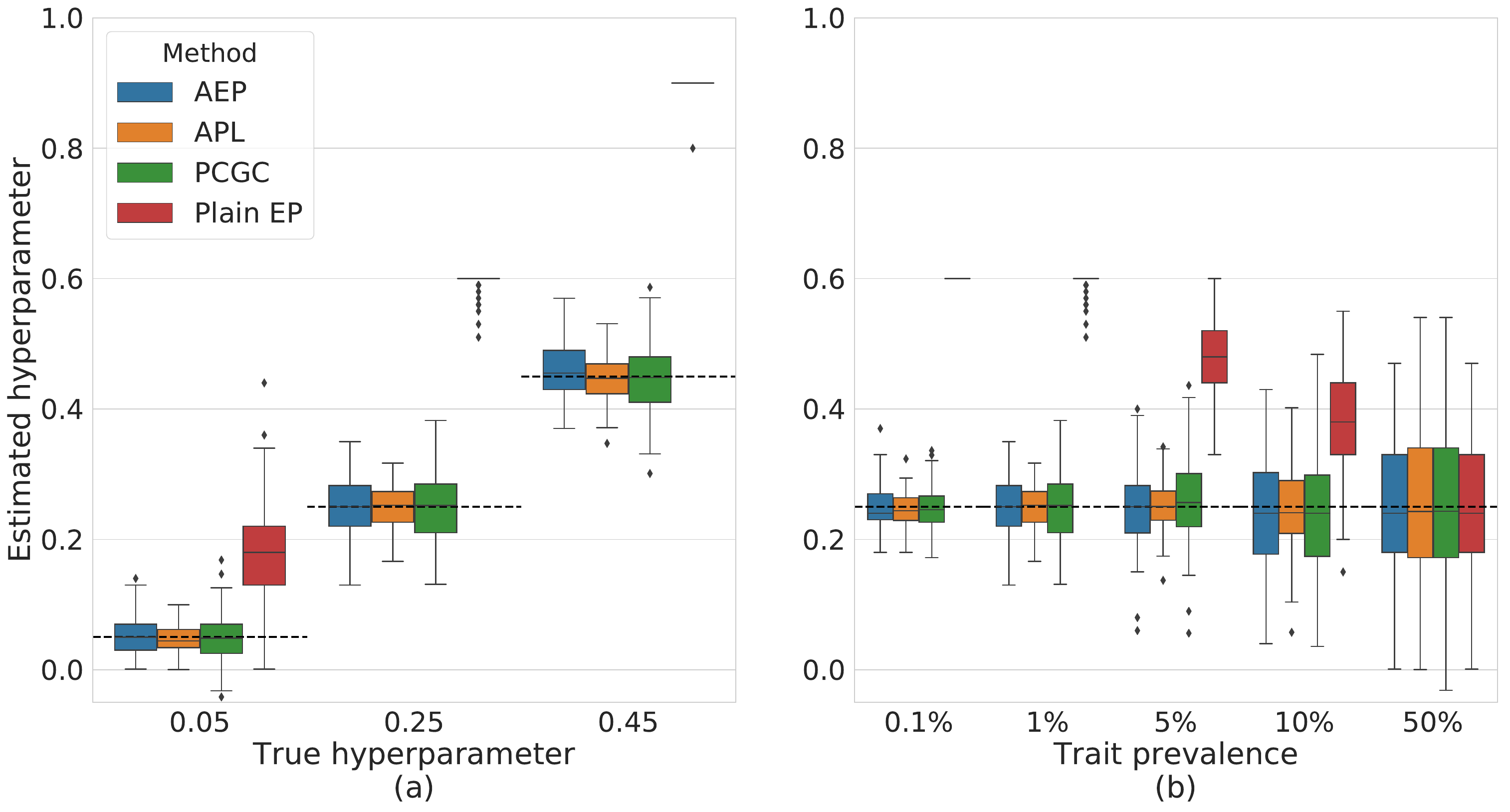}
\caption{\label{fig:fig_var_comp}
Evaluating hyperparameter estimation accuracy. Shown are box-plots depicting the estimates of each method across 100 different simulations, under data sets with an equal number of cases and controls, and a model with a single scale hyperparameter $\sigma^2_g$. The dashed horizontal lines represent the true underlying values of $\sigma^2_g$ used to generate the data.  (a) AEP, APL and PCGC provide accurate estimates of $\sigma^2_g$ when the true trait prevalence (the prevalence of cases in the population) is 1\%, for various values of $\sigma^2_g$, whereas plain EP is severely biased.  (b) All methods except for plain EP accurately estimate $\sigma^2_g$ regardless of the underlying trait prevalence. Plain EP is accurate only when the prevalence is 50\%, in which case there is no ascertainment.
}
\end{figure}

\subsection{Simulation Studies: Estimating hyperparameters}
Our first experiment evaluated variance component estimation accuracy. All methods except plain EP yielded
empirically unbiased estimates, whereas plain EP was severely biased
(Figure \ref{fig:fig_var_comp}a). We also
generated data under different prevalence values \(K\)
and verified that all methods except plain EP remained accurate regardless of \(K\),
whereas plain EP was only accurate when \(K = 0.5\), in which case there
is no ascertainment (Figure \ref{fig:fig_var_comp}b).

\begin{figure}
\centering
\includegraphics[width=1\linewidth]
{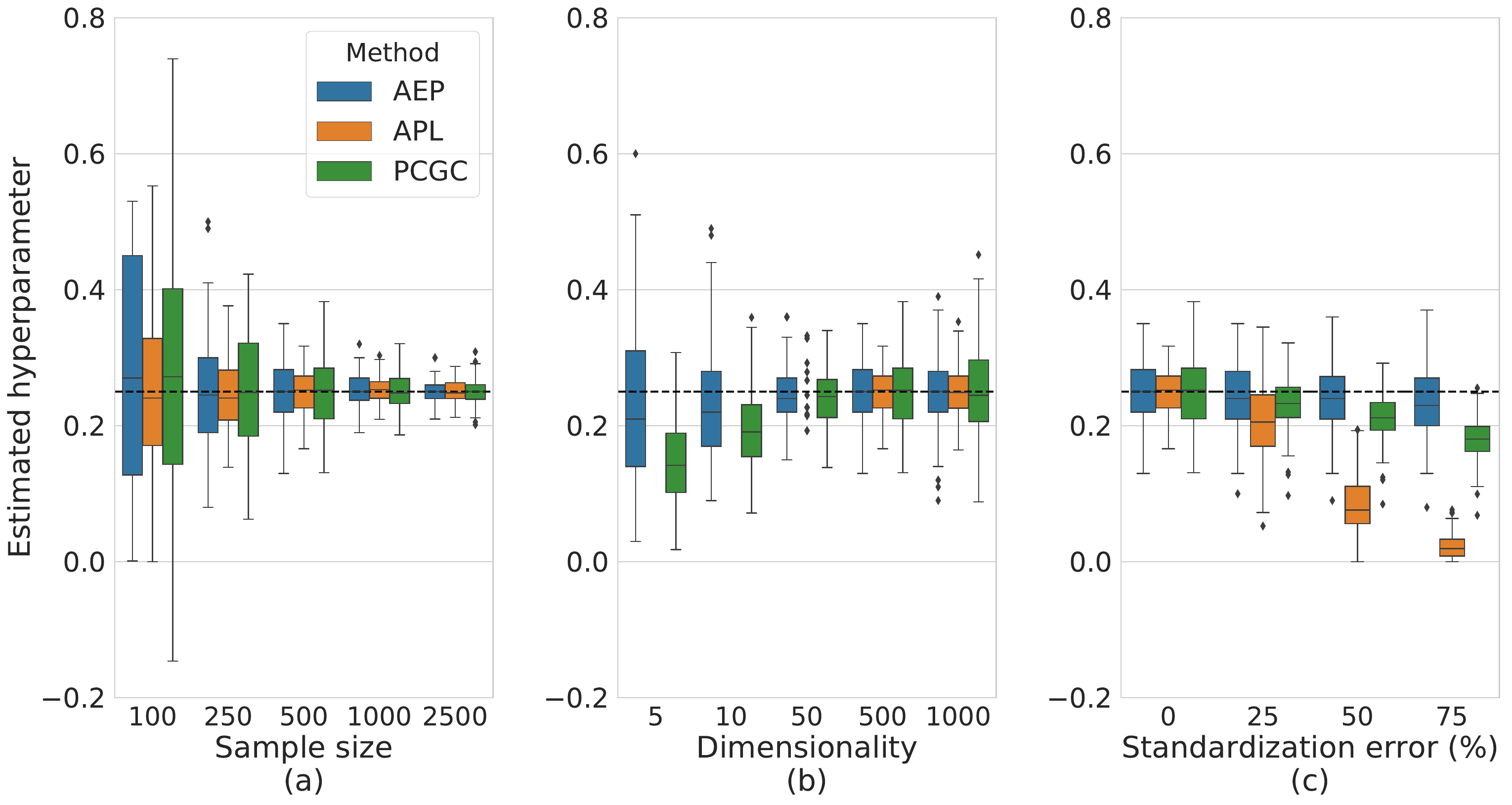}
\caption{\label{fig:fig_simulations}
Investigating how hyperparameter estimation performance is affected by sample size, data dimensionality and modeling violations. (a) All the methods gain accuracy as the sample size increases. The sampling variance of PCGC is consistently larger, because it uses a moment-based rather than a likelihood-based estimator. (b)   All the methods gain accuracy as dimensionality increases. AEP is substantially more accurate than the other methods in the presence of a small number of features, because the other two methods use a first-order Taylor expansion around $\bm{ZZ}^T=\bm{I}$, which is less accurate in the presence of a small number of features. APL estimates for numbers $<$50 are equal to 1.0, and are omitted for clarity. (c) AEP is robust to feature standardization misspecification (see main text), whereas PCGC is moderately sensitive and APL is highly sensitive.}
\end{figure}

In the next experiment we examined sensitivity to sample size $n$ and dimensionality $m$
(corresponding to the number of rows and columns in the matrix
\(\bm{Z}\), respectively). We first verified that all methods became
increasingly accurate with increasing sample size, but PCGC had a
consistently larger sampling variance, because it uses a moment-based rather than a likelihood-based estimator (Figure \ref{fig:fig_simulations}a). We also observed
that all methods became increasingly accurate with increasing dimensionality, but AEP was substantially more accurate when
$m<50$ (Figure \ref{fig:fig_simulations}b). This is because
the other two methods use a first-order Taylor expansion around
\(\bm{Z}\bm{Z}^{T} = \bm{I}\), which is less accurate when $m$ is small.

We also examined robustness to modeling
violations by introducing noise into the
feature standardization procedure. We multiplied
the estimated frequency of every binary variable \(j\) by
\(r_{j} \sim U\left( \frac{1}{1 + e},1 + e \right)\) before
standardizing it, where \(e \in \left\lbrack 0,1 \right\rbrack\) is the
 error magnitude, and used this value for estimation, but
not for the true generative model. This noise
model is motivated by GWAS, where genetic variants are often
standardized according to (somewhat noisy) estimates of their population frequency rather than their sample frequency,
to prevent bias due to ascertainment. AEP was
highly robust to such modeling misspecification, whereas PCGC was
moderately sensitive and APL was highly sensitive to such
misspecification (Figure \ref{fig:fig_simulations}c). These experiments indicate that AEP is
more reliable than the other methods under a wide variety of modeling
assumptions, and is thus the method of choice for GP likelihood-based
hyperparameter estimation in case-control studies.

Next, we examined estimation accuracy under non-linear kernels. We generated data with a scaled RBF kernel, 
$
K_{ij} = 
\sigma^2_g
\text{exp}\left(
-\left\Vert 
\bm{Z}_i - \bm{Z}_j
\right\Vert
^2 
/
\left( 2 \gamma^2 \right)
\right)
$,
and estimated the hyperparameters
$\bm{\theta} = \left\{\sigma^2_g, \gamma \right\}$.
Our generative model used $\sigma^2_g=0.25$, $\gamma=0.5$, and the same values as in the linear kernel simulations for all other parameters, except for restricting to $m=$10 normally-distributed features. This is because RBF kernels tend to overfit under a large $m$, yielding $K_{ij}$ that is very close to either 0 or $\sigma^2_g$ regardless of $\bm{Z}$.

A technical challenge of the RBF experiments is that our simulations first generate true latent variables $g_i$ for a population of 1M units. This requires computing a $1M \times 1M$ RBF covariance matrix $\bm{K}$ and  sampling $\bm{g} = \left\lbrack g_1, \ldots, g_n \right\rbrack^T$ from $\mathcal{N}\left(\bm{0}, \bm{K}\right)$, which is computationally intractable under a non-linear kernel.
Instead we (1) generated a base population of 10K feature vectors $\bm{Z}_i$ and a corresponding $10K \times 10K$ RBF kernel matrix $\bm{K}_{10K}$; (2) sampled 10K $g_i$ values from $\mathcal{N}\left(\bm{0}, \bm{K}_{10K}\right)$; and (3) created a population of 1M units, such that each unit has a vector $\bm{Z}_i$ and a corresponding $g_i$ value selected at random from the base population, along with uniquely generated values of $\bm{X}_i$ and $\epsilon_i$. Afterwards we followed the same procedure as in the linear kernel simulations of (1) generating a liability $l_{i} = \bm{X}_{i}^{T}\bm{\beta} + g_i + \epsilon_{i}$ for each unit; (2) determining the liability cutoff according to the desired prevalence; and (3) sampling $n/2$ cases and $n/2$ controls. We modified PCGC and APL to ignore pairs of units with identical features in these experiments.


AEP was empirically unbiased in the RBF experiments, having an average estimation bias of -0.016 (stdev 0.068) for $\sigma^2_g$ and of 0.022 (stdev 0.076) for $\gamma$ across 100 simulations. In contrast, APL, PCGC and plain EP were severely biased, with
an average bias $>$0.2 in the estimation of $\sigma^2_g$ and $>$0.15 (in absolute value) in the estimation of $\gamma$. We verified that the bias was not due to the modified data generation scheme by repeating the same experiments with a linear kernel, wherein PCGC and APL were empirically unbiased. These results likely arise because PCGC and APL both use a first-order Taylor expansion around $K_{ij}=0$, which may be less accurate in the presence of non-linear kernels. We conclude that PCGC and APL cannot be trivially modified to handle non-linear kernels, whereas AEP can be used in more general settings.

Finally, we examined the computational speed of the methods.
PCGC and APL are very efficient compared
to AEP, because they scale quadratically with the sample size whereas
AEP scales cubically, like standard EP \citep{nickisch_approximations_2008}. Nevertheless, AEP can
 perform maximum likelihood estimation in data sets with 3,000
units in less than two hours, and is thus applicable to solve reasonably
sized real-world problems.
AEP can potentially be scaled up using novel methods developed for GPs (see Discussion).

\begin{figure}
\centering
\includegraphics[width=1\linewidth]{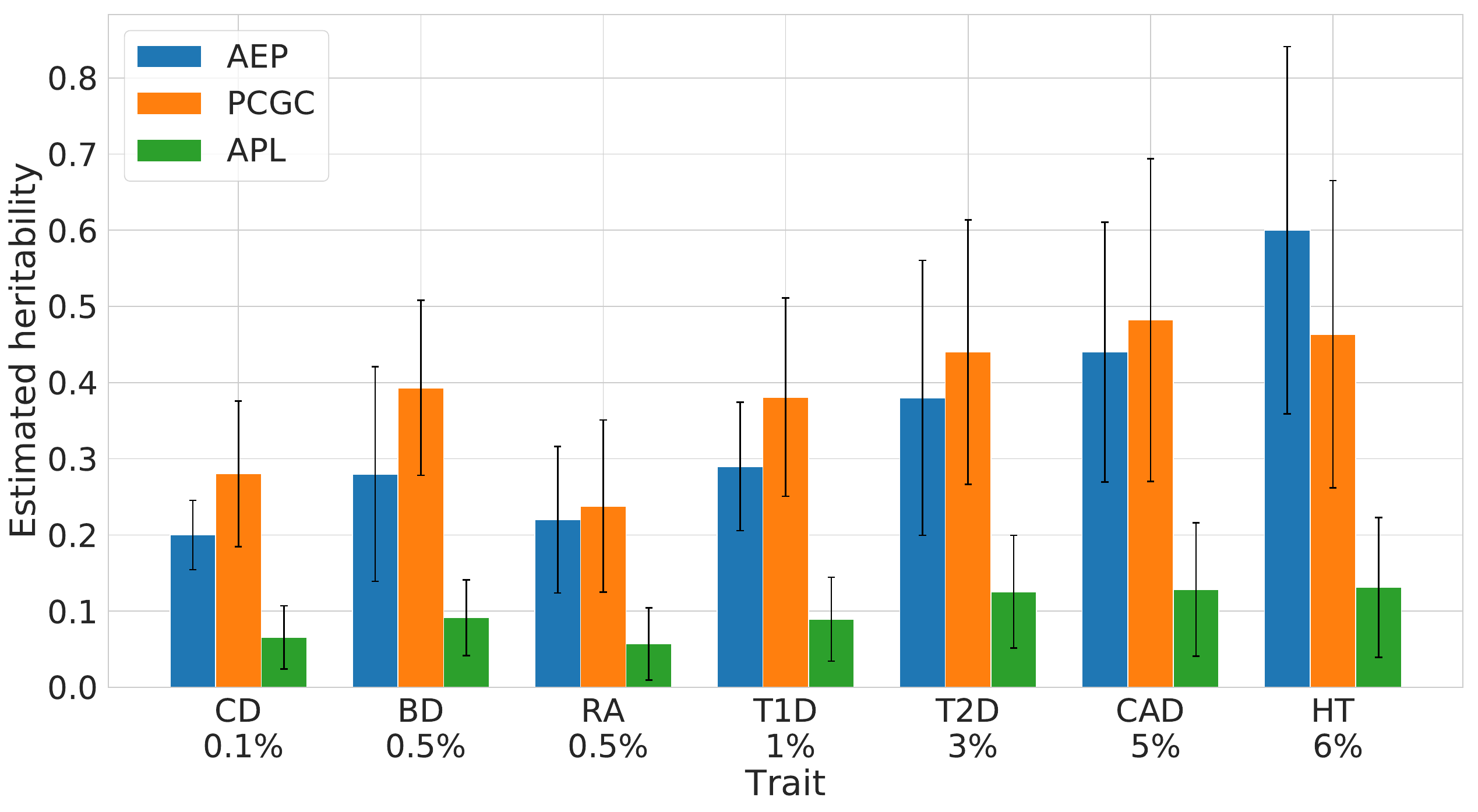}
\caption{\label{fig:WTCCC}
Shown are estimates of heritability (the proportion of liability variance explained by genetic factors) of seven complex disorders from \citep{wtccc_genome-wide_2007}. The error bars are the standard deviation multiplied by 1.96, as estimated via jackknife. The disorders are Crohn’s disease (CD), rheumatoid arthritis (RA), bipolar disorder (BD), type 1 diabetes (T1D), type 2 diabetes (T2D), coronary artery disease (CAD) and hypertension (HT).  The population prevalence of each trait is shown below its name. The estimates of AEP and PCGC are relatively concordant, whereas the APL estimates are significantly down-biased, in agreement with the modeling misspecification simulations.
}
\end{figure}

\subsection{Real Data Analysis}
We evaluated the ability of the proposed methods to estimate GP hyperparameters in real data sets. To this end, we estimated the heritability of seven complex disorders, having population prevalence between 0.1\% and 6\%, based on large data sets with $\sim$3,700 individuals and $\sim$280,000 genetic variants from the Wellcome Trust 1 case-control consortium \citep{wtccc_genome-wide_2007}. We performed stringent preprocessing to avoid confounding artifacts, as reported in our previous publication \citep{weissbrod_multikernel_2016}. We modeled sex, which is strongly associated with several of these traits, as a binary feature associated with a fixed effect, and estimated its effect via AGEE as done in the simulation studies. Standard errors were computed via jackknife sampling.

The heritability explained by measured genotypes for the investigated disorders lies in the range 20\%-60\% (Figure \ref{fig:WTCCC}). There is a high degree of concordance between PCGC and AEP, whereas the APL estimates are substantially lower. This behavior is consistent with the simulation studies and suggests that APL is highly sensitive to model misspecification. AEP is more accurate than the state of the art (PCGC) under simulations and yields similar estimates with a smaller sampling variance in real data analysis, and is thus the first robust scalable method we are aware of for likelihood-based inference in GPs with unit-level ascertainment and a non-sparse, full-rank covariance structure.

\begin{figure}
\centering
\includegraphics[width=1\linewidth]{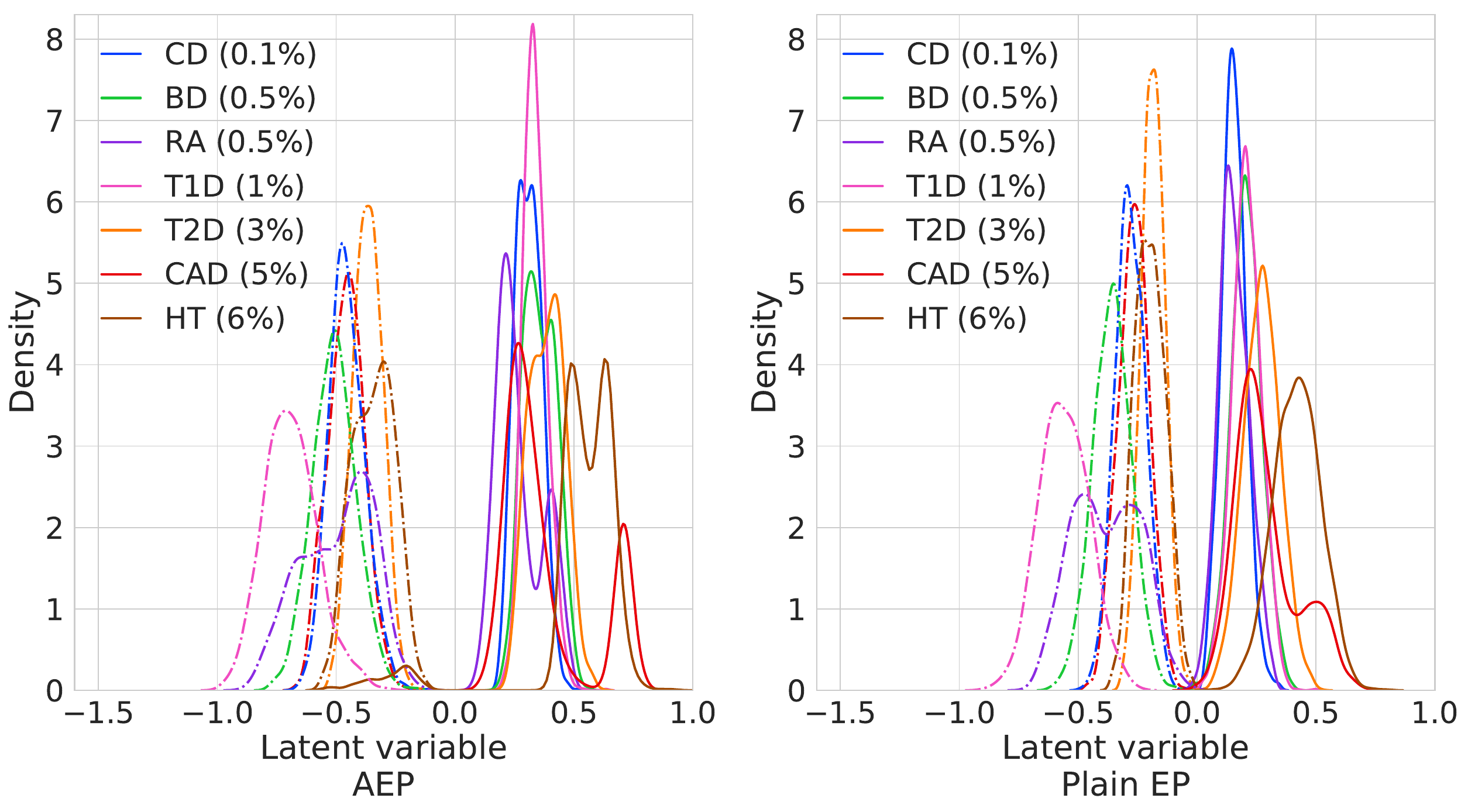}
\caption{\label{fig:posterior_mean}
Inference of latent variables of complex genetic disorders. Shown is the distribution of the posterior mean of latent variables $g_i$ provided by AEP (left) and plain EP (right),  estimated via Gaussian kernel density estimation separately for controls (dashed lines) and cases (solid lines).
The trait names are the same as in Figure \ref{fig:WTCCC}, and prevalences are shown in parentheses. Individuals with larger posterior mean estimates carry a greater genetic load of disease-inducing variants.
AEP provides a clearer separation between cases and controls by exploiting knowledge about the prevalence and sampling scheme.
For both traits, the distribution variance increases with heritability and prevalence.
Note that these distributions are not analogous to the ones in Figure \ref{fig:fig_densities} because they are based on posterior rather than marginal prior distributions.
}
\end{figure}

To further demonstrate the capabilities of AEP we estimated the posterior distribution of the GP latent variables $g_i$ under both AEP and plain EP, using the standard EP approximation of the posterior distribution of latent variables \citep{rasmussen_gaussian_2006}. Inferring the latent variables has a clinical utility as a measure of disease severity: individuals with larger values have a stronger genetic load of disease-inducing variants. The latent variables follow a clear bimodal distribution in all cases, but AEP provides a stronger separation between cases and controls (Figure \ref{fig:posterior_mean}). This result cannot be obtained with existing methods, because no existing likelihood-based GP approximation can model ascertainment.

\section{Discussion}
We presented several methods for inference of GP hyperparameters
in settings with unit-level ascertainment, and a full-rank, non-sparse covariance structure. This was done by
combining the ascertained likelihood framework with GPs and GLMMs, which form the statistical backbone of likelihood based analysis of non-iid data.

We proposed two approximate likelihood-based methods for the ascertained
GP framework, AEP and APL, and empirically compared them with PCGC---the current state of the art method for estimating variance components in genetic case-control studies---which uses a moment-based rather than a
likelihood-based estimator. APL is very computationally efficient but
is sensitive to model misspecification. AEP, which is
the most complex and empirically best approximation of maximum likelihood we
proposed, is slower and is more technically complex than APL and PCGC, but is consistently more accurate than PCGC, and is less
sensitive to modeling assumptions in our simulations. AEP additionally
has the advantage of providing a full probabilistic model with a
well-defined likelihood, and it recovers standard EP as a special case
under random ascertainment. On the other hand, PCGC has a principled
underlying approximation, whereas APL and AEP are less well understood.
Hence, the three methods are complementary in terms of
their strengths and weaknesses, and we encourage future case-control
studies to use multiple methods to gain a deeper understanding of high
dimensional dependency structures.

The combination of unit-level ascertainment and a full-rank, non-sparse covariance structure is very common in statistical
genetics \citep{golan_measuring_2014}, but is often encountered in other scientific domains,
such as geostatistics and GP classification \citep{diggle_model-based_1998,chu_classification_2010,ziegler_individualized_2014,young_accurate_2013}.
Ascertained sampling is almost inevitable when studying rare phenomena,
and the increasing dimensionality of studied data often necessitates the
introduction of random rather than fixed effects, which in turn induce
full-rank, non-sparse dependency structures. Additionally, it is often more convenient
to perform dense sampling in a small number of clusters rather than
collecting a large number of clusters \citep{bellamy_quantifying_2005,zhang_inconsistent_2004,glidden_modelling_2004}, leading to non-sparse,
full-rank dependency structures at the cluster level. Hence, we expect
our work to be applicable in diverse scientific fields.

In this study we extended the well-known EP algorithm \citep{minka_expectation_2001} to approximate
 GP likelihood. Another potential approach is MCMC
sampling coupled with an integration scheme such as thermodynamic
integration \citep{kuss_assessing_2005,nickisch_approximations_2008,gelman_simulating_1998}, but in our experience such approaches are too slow and complex for modern sized data sets. In recent years, Bayesian approaches have proven to be potential
alternatives to likelihood based approaches in GPs \citep{ferkingstad_improving_2015}. However, such
approaches can be sensitive to the choice of prior distribution, and
require prohibitively computationally expensive MCMC sampling.
Several analytical approximations exist, but these are often 
inaccurate in the presence of binary data \citep{fong_bayesian_2010}. The potential use
of sampling-based approaches for inference in GPs under case-control
ascertainment remains to be explored.

In recent years, genetic biobanks with hundreds of thousands of individuals have become available \citep{bycroft2018uk}. AEP scales cubically with sample size and is thus not scalable to such  datasets. GP approximation techniques from the machine learning  community, such as mixture-of-experts models \citep{pmlr-v37-deisenroth15}, inducing points \citep{snelson2006sparse,wilson2015kernel,pmlr-v84-gardner18a}, random feature expansions \citep{rahimi2008random,le_fastfood_2013,yang_cartelearning_2015}  and stochastic variational approximations \citep{hensman_gaussian_2013,wilson2016stochastic,cheng2017variational}, or from the GWAS community (e.g. \citealt{loh_contrasting_2015}) can potentially be used to scale up AEP to such large datasets.

A major challenge of non-sparse dependencies is that
statistical theory is relatively undeveloped for this case.
Specifically, assuming a study with \(r\) mutually independent clusters
of \(m\) units, statistical theory is well developed for the asymptotic
behavior \(r/m \rightarrow \infty\), but is limited for
\(r/m \rightarrow 0\), which is our setting of interest (as
\(r\)\emph{=1} when the covariance matrix is non-sparse). The statistical consistency of
estimators in such cases has been established in limited settings,
including GEEs \citep{xie_asymptotics_2003}, maximum penalized quasi likelihood \citep{bellamy_quantifying_2005}, composite
likelihood approximation \citep{heagerty_composite_1998}, Laplace approximations \citep{shun_laplace_1995}, and specific
geostatistical models \citep{zhang_towards_2005,du_fixed-domain_2009}. Several studies have established
the statistical consistency of maximum likelihood estimators for linear mixed models (i.e., GPs with a linear kernel and a normal likelihood) in similar settings using random matrix theory \citep{bonnet_heritability_2015,jiang_high-dimensional_2016,dicker_maximum_2016}, but to our
knowledge such results have not been derived for GPs with non-normal likelihoods. We conclude that
there is a major gap in statistical theory regarding
\(r/m \rightarrow 0\) asymptotics, representing questions of both
theoretical and practical importance.

Several topics that remain unexplored in this work are GPs
with more advanced kernels, outcome prediction and testing of
fixed effects, for which several heuristic methods have been proposed in
the statistical genetics literature \citep{hayeck_mixed_2015,weissbrod_accurate_2015,chen_control_2016,jiang_retrospective_2016}. Extending our approach to
handle these topics is a potential avenue for future work.


\acks{This work was supported by grant 1804/16 from the Israel Science Foundation. This study makes use of data generated by the Wellcome Trust Case Control Consortium. A full list of the investigators who contributed to the generation of the data is available from www.wtccc.org.uk. Funding for the project was provided by the Wellcome Trust under award 076113. We thank Malka Gorfine for fruitful discussions.}

\appendix

\section*{Appendix A}
Here we describe a fast Taylor expansion-based approximation to APL estimation of GPs with a probit likelihood and a scaling parameter $\sigma^2$. Denote $\textrm{cov}(g_i,g_j)=\rho \sigma^2_g$, where $g_i$ is the latent variable of unit $i$ and $\rho$ depends on $\bm{Z}_i$,$\bm{Z}_j$,  and on all the other kernel hyperparameters. The joint likelihood of each pair of units can be written as:
\begin{align}
P(y_i=a,y_j=b \,|\, \bm{X}_i, \bm{X}_j, \bm{Z}_i, \bm{Z}_j, s_i=s_j=1)
& = \frac{A_{ab}(\rho)} {B(\rho)}
P(s_i=1 \,|\, y_i)
P(s_j=1 \,|\, y_j), \label{eq:PL_bayes2}
\end{align}
where $
A_{ab}(\rho)  \triangleq P(y_i=a,y_j=b \,|\, \bm{X}_{i},\bm{X}_{j}, \rho)
$,
$
B(\rho) \triangleq P(s_i=s_j=1\,|\,\bm{X}_{i},\bm{X}_{j}, \rho)
$, and we omitted the hyperparameters $\bm{\theta}$, $\bm{\beta}$ for brevity.
Using the law of total probability, we can write:
$
B(\rho) = (s^1)^2A_{11}(\rho) + s^1s^0\left(A_{10}(\rho)+A_{01}(\rho)\right) + (s^0)^2 A_{00}(\rho)
$,
where $s^t = P(s_i=1 | y_i=t)$.
Next, we explicitly evaluate these quantities at $\rho=0$:
$
A_{ab}(0) = K_i^a (1-K_i)^{1-a} K_j^b (1-K_j)^{1-b}
$,
$
B(0) = \left(s^0(1-K_i) + s^1K_i\right)  \left(s^0(1-K_j) + s^1K_j\right)
$,
where $K_i = P(y_i=1 | \bm{X}_i)$, and we omitted the dependence on $\bm{Z}_i$ because we assume that 
$g_i \sim \mathcal{N}(0, \sigma^2_g)$ marginally
regardless of $\bm{Z}_i$.
The above equations hold because $y_i,y_j$ and $s_i,s_j$ are independent given $\rho=0$.

We next compute the partial derivatives of both expressions with respect to $\rho$ at $\rho=0$. Following \citep{golan_measuring_2014}, we have:
\begin{align}
\frac{d}{d\rho}A_{ab}(\rho) | _{\rho=0} & = \phi(t_i) \phi(t_j)\sigma^2 (-1)^{a\neq b}\nonumber  \\
\frac{d}{d\rho}B(\rho) | _{\rho=0} & = (s^1)^2 \frac{d}{d\rho}A_{11}(\rho) | _{\rho=0} + 2s^1s^0 \frac{d}{d\rho}A_{a \neq b}(\rho) | _{\rho=0} + (s^0)^2 \frac{d}{d\rho}A_{00}(\rho) | _{\rho=0} \nonumber \\
& = \phi(t_i) \phi(t_j)\sigma^2 \left((s^1)^2+(s^0)^2 - 2s^1s^0\right), \nonumber
\end{align}
where $\phi(\cdot)$ is the standard normal density, and $t_i = \Phi^{-1}(1-K) - \bm{X}_i^T\bm{\beta}$ is the liability cutoff for unit $i$, with $\Phi(\cdot)$ representing the standard normal cumulative density and $K$ being the prevalence of cases in the population.

Finally, we plug in the above expressions into the Taylor expansion of Equation \ref{eq:PL_bayes2} at $\rho=0$, which can be written as follows:
\begin{align}
\frac
{A_{ab}(\rho)}{B(\rho)}
P(s_i \,|\, y_i)
P(s_j \,|\, y_j) =
& \left(
\frac{A'_{ab}(0)B(0) - B'(0)A_{ab}(0)}{B(0)^2} \rho + \mathcal{O}(\rho^2) \nonumber
\right) P(s_i \,|\, y_i)
P(s_j \,|\, y_j).
\end{align}

\section*{Appendix B}

Here we provide an informal analysis motivating the use of AEP. A formal analysis of AEP is difficult because there are relatively few theoretical guarantees for standard EP \citep{dehaene_bounding_2016, dehaene2018expectation}. Instead, we state several assumptions and then provide an informal analysis under these assumptions.
Throughout this Appendix, the notation $\bm{s}$ is a shorthand notation for $s_1= \ldots=s_n=1$, $\bm{u}_{-i}$ indicates the vector $\bm{u}$ with the $i^{\text{th}}$ entry removed, and we omit the dependence on \(\bm{\beta},\bm{\theta}\) for brevity.

\subsection*{The parametric form of the AEP site approximation \label{sec:AEP_consistency}}

We first demonstrate that in order for AEP to approximate the ascertained likelihood $P\left(\bm{y} | \bm{X},\bm{Z},\bm{s}\right)$ (up to a scaling factor) at its fixed point, the site function $t_i(g_i)$ needs to approximately take the following parametric form:
\begin{align}
t_i(g_i) \approx \frac{P(y_i,s_i|g_i,\bm{X}_i)}{P(s_i | \bm{X},\bm{Z},\bm{s}_{-i})}. \label{eq:t_ig_i}
\end{align}
Note that this is different from the standard EP approximation $t_i(g_i) \approx P(y_i | X_i, g_i)$.

Our derivation requires an additional approximation:
\begin{approximation}
\label{Approx1}
\textup{
\begin{align}
P(\bm{s} | \bm{X},\bm{Z}) \approx \frac{1}{C(\bm{X}, \bm{Z})} \prod_{i} P(s_i | \bm{X},\bm{Z},\bm{s}_{-i}),
\hspace{10pt}
\text{for some proportionality factor} \hspace{5pt} C(\bm{X}, \bm{Z}).
 \nonumber
\end{align}
}
\end{approximation}
Approximation \ref{Approx1} is motivated by the theory of composite likelihood estimators \citep{varin_overview_2011}. Specifically, the composite maximum likelihood estimator of \(\bm{\beta},\bm{\theta}\) is asymptotically normally distributed around their true values under suitable regularity conditions \citep{cox2004note}. Since both the full and the composite maximum likelihood estimators are asymptotically normal with the same mean, the composite likelihood is approximately proportional to the full likelihood around this mean, with the approximation accuracy depending on the ratio between their variances. This ratio depends on the ratio between the diagonal entries of the Fisher and the Godambe information matrices \citep{varin_overview_2011}.
Note that if Approximation \ref{Approx1} holds, this implies that it also holds when replacing $\bm{s}$ with $\bm{s}_{-j}$ and omitting $j$ from the product. This property will be used in the derivations below.

Under Equation \ref{eq:t_ig_i} and Approximation \ref{Approx1}, the AEP likelihood
$\int P(\bm{g}|\bm{Z}) \prod_i t_i(g_i)d\bm{g}$
is approximately proportional to the ascertained likelihood $P\left(\bm{y} | \bm{X},\bm{Z},\bm{s}\right)$:
\begin{align}
\int P(\bm{g}|\bm{Z}) \prod_i t_i(g_i)d\bm{g} 
& \stackrel{\text{Equation \ref{eq:t_ig_i}}}{\approx} \int P(\bm{g}|\bm{Z}) \prod_i  \frac{P(y_i,s_i | g_i,\bm{X}_i)} {P(s_i | \bm{X},\bm{Z},\bm{s}_{-i})} d\bm{g} \nonumber \\
& = \int \frac{ P(\bm{g}|\bm{Z}) \prod_i P(s_i | g_i,\bm{X}_i)}{\prod_i P(s_i | \bm{X},\bm{Z},\bm{s}_{-i})} \prod_i P(y_i | s_i,g_i,\bm{X}_i)d\bm{g} \nonumber \\
& \stackrel{\text{Approximation \ref{Approx1}}}{\approx} C(\bm{X}, \bm{Z}) \int \frac{ P(\bm{g}|\bm{Z}) P(\bm{s}|\bm{g},\bm{X})}{P(\bm{s}|\bm{X},\bm{Z})} P(\bm{y}|\bm{s},\bm{g},\bm{X}) d\bm{g} \nonumber \\
& =C(\bm{X}, \bm{Z})  \int P(\bm{g}|\bm{X},\bm{Z},\bm{s})P(\bm{y}|\bm{s},\bm{g},\bm{X})d\bm{g} \nonumber \\ 
& = C(\bm{X}, \bm{Z}) \cdot P(\bm{y}|\bm{X},\bm{Z},\bm{s}). \nonumber
\end{align}
The last two equalities use the fact that $\bm{y}$, $\bm{s}$ are conditionally independent of $\bm{Z}$ given $\bm{g}$.
We conclude that if $t_i(g_i)$ approximately takes the form of Equation \ref{eq:t_ig_i} then the hyperparameters $\bm{\beta}$, $\bm{\theta}$ which maximize the AEP likelihood are approximately the maximum likelihood estimates.

\subsection*{Derivation of the AEP Step \label{sec:AEP_step_deriv}}

Here we provide a heuristic motivation for the AEP step procedure. Recall from Section \ref{subsec:EP} that the AEP step 
consists of finding the unnormalized Gaussian $t_i(g_i)$ that optimizes the following approximation:
\begin{align}
\int q_{-i}(g_i)t_i(g_i) dg_i
\approx
\frac{\int q_{-i}(g_i)P(y_i,s_i|g_i,\bm{X}_i)dg_i}{\int q_{-i}(g_i)P(s_i|g_i,\bm{X}_i)dg_i}, \label{eq:AEP_step}
\end{align}
where $q_{-i}(g_i) \propto \int{P(\bm{g}|\bm{Z}) \prod_{j \neq i}t_j(g_j) d\bm{g}_{-i}}$, and the optimization is performed by matching the zeroth, first, and second derivatives of both functions with respect to $\mu_{-i}$.

We first write down the natural analogue of the standard EP step objective for AEP.
According to Equation \ref{eq:t_ig_i}, this objective finds the unnormalized Gaussian $t_i(g_i)$ that optimizes the approximation:
\begin{align}
\int q_{-i}(g_i)t_i(g_i) dg_i \approx \int q_{-i}(g_i) \frac{P(y_i,s_i|g_i,\bm{X}_i)}{P(s_i | \bm{X},\bm{Z},\bm{s}_{-i})} dg_i. \label{eq:EP_asc_goal_best_nostar}
\end{align}
Unlike standard EP, we cannot minimize the GKL divergence between the functions in the integrals in Equation \ref{eq:EP_asc_goal_best_nostar}, because this will lead to the same solution $t_i(g_i)$ as in standard EP up to a scaling factor. To see this, note that the function in the rhs of Equation \ref{eq:EP_asc_goal_best_nostar} can be written as $q_{-i}(g_i)P(y_i | g_i, \bm{X}_i) W$, where $W = \frac{P(s_i | y_i)}{P(s_i | \bm{X},\bm{Z},\bm{s}_{-i})}$ is constant with respect to $g_i$. Hence, minimizing the GKL divergence will lead to the same approximation as in standard EP, up to the scaling factor $W$.

Instead of minimizing the GKL divergence, we will approximate the rhs of Equation \ref{eq:EP_asc_goal_best_nostar} as follows:
\begin{align}
\int q_{-i}(g_i) \frac{P(y_i,s_i|g_i,\bm{X}_i)}{P(s_i | \bm{X},\bm{Z},\bm{s}_{-i})} dg_i
\approx
\frac{\int q_{-i}(g_i)P(y_i,s_i|g_i,\bm{X}_i)dg_i}{\int q_{-i}(g_i)P(s_i|g_i,\bm{X}_i)dg_i} \label{eq:rhs_approx_main}.
\end{align}
The AEP step in Equation \ref{eq:AEP_step} is obtained by equating the lhs of Equation \ref{eq:EP_asc_goal_best_nostar} with the rhs of Equation \ref{eq:rhs_approx_main}.

It remain to derive Equation \ref{eq:rhs_approx_main}. Our derivation uses the following assumption:
\begin{assumption}
\label{assumption1}
Weak dependence between \textup{$\bm{y}_{-i}$} and \textup{$s_i$} conditional on \textup{$\bm{X}$} and on \textup{$\bm{Z}$}:
\normalfont
\begin{align}
P(\bm{y}_{-i} | \bm{X},\bm{Z},\bm{s}) \approx P(\bm{y}_{-i} | \bm{X},\bm{Z},\bm{s}_{-i}). \nonumber
\end{align}
\end{assumption}
The derivation additionally uses the following two approximations, which we derive below by using Equation \ref{eq:t_ig_i}, Approximation \ref{Approx1} and Assumption \ref{assumption1}:
\begin{approximation}
\label{Approx2}
At the fixed point we have: \normalfont
$
q_{-i}(g_i) \approx P(g_i, \bm{X},\bm{Z}, \bm{y}_{-i}). \nonumber
$
\end{approximation}
\begin{approximation}
\label{Approx3}
\normalfont
$
\int q_{-i}(g_i) \frac{P(y_i,s_i|g_i,\bm{X}_i)}{P(s_i | \bm{X},\bm{Z},\bm{s}_{-i})} dg_i
\approx P(y_i | \bm{X},\bm{Z},\bm{y}_{-i},s_i). \nonumber
$
\end{approximation}
We complete the derivation of Equation \ref{eq:rhs_approx_main} by first using Approximation \ref{Approx3} to approximate the rhs of Equation \ref{eq:EP_asc_goal_best_nostar} and the lhs of Equation \ref{eq:rhs_approx_main} as $P(y_i | \bm{X},\bm{Z},\bm{y}_{-i},s_i)$, and then using Approximation \ref{Approx2} and the graphical model structure (Figure \ref{fig:graphical_model}) to approximate $P(y_i | \bm{X},\bm{Z},\bm{y}_{-i},s_i)$ via $q_{-i}(g_i)$ as follows:
\begin{align}
P(y_i | \bm{X},\bm{Z},\bm{y}_{-i},s_i) &
= \frac
{P(y_i,s_i | \bm{X},\bm{Z},\bm{y}_{-i})}
{P(s_i | \bm{X},\bm{Z},\bm{y}_{-i})} \nonumber \\
& =
\frac
{\int P(g_i | \bm{X},\bm{Z},\bm{y}_{-i}) P(y_i,s_i | \bm{X}_i,g_i) dg_i}
{\int P(g_i | \bm{X},\bm{Z},\bm{y}_{-i}) P(s_i | \bm{X}_i,g_i) dg_i}
\nonumber \\
& \approx 
\frac
{\int q_{-i}(g_i) P(y_i,s_i | \bm{X}_i,g_i) dg_i}
{\int q_{-i}(g_i) P(s_i | \bm{X}_i,g_i) dg_i}.
\nonumber
\end{align}
This completes the derivation.

\subsection*{Derivation of  Approximations  \ref{Approx2}--\ref{Approx3}}

We now provide heuristic derivations of Approximations  \ref{Approx2}--\ref{Approx3}.
\setcounter{approximation}{1}
\begin{approximation}
\normalfont
$q_{-i}(g_i) \approx P(g_i, \bm{X},\bm{Z}, \bm{y}_{-i}).$
\end{approximation}
\normalfont
Our derivation consists of two stages. First, we define the unnormalized cavity distribution $q^{*}_{-i}(g_i) \triangleq \int{P(\bm{g}|\bm{Z}) \prod_{j \neq i}t_j(g_j) d\bm{g}_{-i}}$, and show that $q^{*}_{-i}(g_i) \approx C(\bm{X}, \bm{Z}) \cdot P(g_i, \bm{y}_{-i} | \bm{X},\bm{Z}, \bm{s}_{-i})$:
\begin{align}
q^{*}_{-i}(g_i) & \triangleq \int P(\bm{g}|\bm{Z}) \prod_{j \neq i}t_j(g_j) d\bm{g}_{-i}  \nonumber \\ 
& \hspace{-25pt} \stackrel{\text{Equation \ref{eq:t_ig_i}}}{\approx} \int P(\bm{g}|\bm{Z}) \prod_{j \neq i} \frac{P(y_j,s_j|g_j,\bm{X}_j)}{P(s_j | \bm{X},\bm{Z},\bm{s}_{-j})} d\bm{g}_{-i} \nonumber \\
& \hspace{-25pt} \stackrel{\text{Approximation \ref{Approx1}}}{\approx} C(\bm{X}, \bm{Z})  \int \frac{P(\bm{g}|\bm{Z})}{P(\bm{s}_{-i} | \bm{X},\bm{Z})} \prod_{j \neq i} P(y_j,s_j|g_i,\bm{X}_j) d\bm{g}_{-i} \nonumber  \\
& \hspace{-25pt}= C(\bm{X}, \bm{Z}) \int \frac{P(\bm{g}_{-i}|\bm{Z}) P(g_i|\bm{g}_{-i},\bm{Z})} {P(\bm{s}_{-i} | \bm{X},\bm{Z})} P(\bm{y}_{-i}, \bm{s}_{-i}| \bm{g}_{-i},\bm{X}_{-i}) d\bm{g}_{-i} \nonumber  \\
& \hspace{-25pt} \stackrel{\text{rearrangement}}{=} C(\bm{X}, \bm{Z}) \int \frac{P(\bm{g}_{-i}|\bm{Z}) P(\bm{s}_{-i} | \bm{g}_{-i},\bm{X}_{-i})} {P(\bm{s}_{-i} | \bm{X},\bm{Z})} P(\bm{y}_{-i}, |\bm{g}_{-i},\bm{s}_{-i},\bm{X}_{-i}) P(g_i|\bm{g}_{-i},\bm{Z}) d\bm{g}_{-i} \nonumber  \\
& \hspace{-25pt} \stackrel{\text{Bayes rule}}{=} C(\bm{X}, \bm{Z}) \int P(\bm{g}_{-i}|\bm{X},\bm{Z}, \bm{s}_{-i}) P(\bm{y}_{-i}, |\bm{g}_{-i},\bm{s}_{-i},\bm{X}_{-i}) P(g_i|\bm{g}_{-i}, \bm{Z}) d\bm{g}_{-i} \nonumber  \\
& \hspace{-25pt} \stackrel{\text{graphical model}}{=} C(\bm{X}, \bm{Z}) \int P(\bm{g}_{-i}, \bm{y}_{-i}|\bm{X},\bm{Z}, \bm{s}_{-i}) P(g_i|\bm{g}_{-i}, \bm{X},\bm{Z}, \bm{s}_{-i}) d\bm{g}_{-i} \nonumber  \\
&  \hspace{-25pt}= C(\bm{X}, \bm{Z}) \int P(\bm{g}, \bm{y}_{-i}|\bm{X},\bm{Z}, \bm{s}_{-i})  d\bm{g}_{-i} \nonumber  \\
& \hspace{-25pt} = C(\bm{X}, \bm{Z}) \cdot P(g_i, \bm{y}_{-i} | \bm{X},\bm{Z},\bm{s}_{-i}) \nonumber
\end{align}
Next, we note that since $q_{-i}(g_i) \triangleq \frac{q^{*}_{-i}(g_i)}{\int q^{*}_{-i}(g_i') dg_i'}$ is a normalized distribution over $g_i$, we have $q_{-i}(g_i) \approx P(g_i | \bm{X},\bm{Z},\bm{y}_{-i}, \bm{s}_{-i})$. Finally, we note that $g_i$ is conditionally independent of $\bm{s}_{-i}$ given $\bm{y}_{-i}$ due to the graphical model structure, yielding $q_{-i}(g_i) \approx P(g_i | \bm{X},\bm{Z},\bm{y}_{-i})$.

\begin{approximation}
\normalfont
$
\int q_{-i}(g_i) \frac{P(y_i,s_i|g_i,\bm{X}_i)}{P(s_i | \bm{X},\bm{Z},\bm{s}_{-i})} dg_i
\approx P(y_i | \bm{X},\bm{Z},\bm{y}_{-i},s_i).
$
\end{approximation}

\normalfont
First, we invoke Approximation \ref{Approx2} and the graphical model structure to obtain the following approximation:
\begin{align}
\int q_{-i}(g_i) \frac{P(y_i,s_i|g_i,\bm{X}_i)}{P(s_i | \bm{X},\bm{Z},\bm{s}_{-i})} dg_i
& \stackrel{\text{Approximation \ref{Approx2}}}{\approx} 
\int P(g_i | \bm{X},\bm{Z},\bm{y}_{-i}) \frac{P(y_i,s_i|g_i,\bm{X}_i)}{P(s_i | \bm{X},\bm{Z},\bm{s}_{-i})} dg_i \nonumber \\
& = \int  \frac{P(g_i | \bm{X},\bm{Z}) P(\bm{y}_{-i} | g_i, \bm{X},\bm{Z})}{P(\bm{y}_{-i} | \bm{X},\bm{Z})} \frac{P(y_i,s_i|g_i,\bm{X}_i)}{P(s_i | \bm{X},\bm{Z},\bm{s}_{-i})} dg_i \nonumber \\
& = \int  \frac{P(g_i | \bm{X},\bm{Z}) P(\bm{y}_{-i} | g_i, \bm{X},\bm{Z})}{P(\bm{y}_{-i} | \bm{X},\bm{Z})}
\frac{P(y_i|g_i,\bm{X}, \bm{Z}, \bm{y}_{-i}) P(s_i|y_i)}{P(s_i | \bm{X},\bm{Z},\bm{s}_{-i})} dg_i \nonumber \\
& = \frac{\int   P(g_i | \bm{X},\bm{Z})  P(\bm{y} | g_i, \bm{X},\bm{Z}) dg_i P(s_i|y_i)}
{P(\bm{y}_{-i} | \bm{X},\bm{Z}) P(s_i | \bm{X},\bm{Z},\bm{s}_{-i})}  \nonumber \\
& = \frac{P(\bm{y} | \bm{X},\bm{Z}) P(s_i|y_i)}
{P(\bm{y}_{-i} | \bm{X},\bm{Z}) P(s_i | \bm{X},\bm{Z},\bm{s}_{-i})}.  \label{eq:Approx2_proof1}
\end{align}
Next, we multiply the rhs of Equation \ref{eq:Approx2_proof1} by
$\frac
{P(\bm{s}_{-i}|\bm{y}_{-i}) P(\bm{s}_{-i} | \bm{X}, \bm{Z})}
{P(\bm{s}_{-i}|\bm{y}_{-i}) P(\bm{s}_{-i} | \bm{X}, \bm{Z})}
$ and invoke Assumption \ref{assumption1}:
\begin{align}
\frac{P(\bm{y} | \bm{X},\bm{Z}) P(s_i|y_i)}
{P(\bm{y}_{-i} | \bm{X},\bm{Z}) P(s_i | \bm{X},\bm{Z},\bm{s}_{-i})}
\frac
{P(\bm{s}_{-i}|\bm{y}_{-i}) P(\bm{s}_{-i} | \bm{X}, \bm{Z})}
{P(\bm{s}_{-i}|\bm{y}_{-i}) P(\bm{s}_{-i} | \bm{X}, \bm{Z})} & = 
\frac{P(\bm{y} | \bm{X},\bm{Z})}
{P(\bm{y}_{-i} | \bm{X},\bm{Z})}
\frac
{P(\bm{s} | \bm{y})  P(\bm{s}_{-i} | \bm{X}, \bm{Z})}
{P(\bm{s} | \bm{X}, \bm{Z})   P(\bm{s}_{-i}|\bm{y}_{-i})} \nonumber \\
& \hspace{-200pt} = 
\frac
{P(\bm{y} | \bm{X},\bm{Z})   
\frac
{P(\bm{s} | \bm{y})}
{P(\bm{s} | \bm{X}, \bm{Z})}
}
{P(\bm{y}_{-i} | \bm{X},\bm{Z})
\frac
{P(\bm{s}_{-i}|\bm{y}_{-i})}
{P(\bm{s}_{-i} | \bm{X}, \bm{Z})}
} \nonumber \\
& \hspace{-200pt} = \frac{P(\bm{y} | \bm{X},\bm{Z},\bm{s})} {P(\bm{y}_{-i} | \bm{X},\bm{Z},\bm{s}_{-i})} \nonumber \\
& \hspace{-200pt} = \frac{P(\bm{y}_{-i}|\bm{X},\bm{Z},\bm{s}) P(y_i | \bm{X},\bm{Z},\bm{y}_{-i},\bm{s})}{P(\bm{y}_{-i}|\bm{X},\bm{Z},\bm{s}_{-i})} \nonumber \\
& \hspace{-200pt} \stackrel{\text{Assumption \ref{assumption1}}}{\approx} \frac{P(\bm{y}_{-i}|\bm{X},\bm{Z},\bm{s}_{-i}) P(y_i | \bm{X},\bm{Z},\bm{y}_{-i},\bm{s})}{P(\bm{y}_{-i}|\bm{X},\bm{Z},\bm{s}_{-i})}  \nonumber \\
& \hspace{-200pt} = P(y_i | \bm{X},\bm{Z},\bm{y}_{-i},\bm{s}) \nonumber \\
& \hspace{-200pt} = P(y_i | \bm{X},\bm{Z},\bm{y}_{-i},s_i) \nonumber
\end{align}


\section*{Appendix C}
Here we describe our novel development of ascertained generalized estimating equations (AGEE).
GEEs are extensions of generalized linear models that can estimate fixed effects while accounting for
dependencies without requiring a probabilistic model \citep{liang_regression_1993}. GEEs require
a correct specification of the mean of the outcome conditional on the
features,
\(\mu_{i} = E\left\lbrack y_{i}|\bm{X}_{i},\bm{\beta} \right\rbrack\),
and a (possibly misspecified) working covariance matrix of the outcomes,
denoted as \(\bm{\Omega}\left( \theta^{\Omega} \right)\) and
parameterized by \(\theta^{\Omega}.\) Given these, \(\bm{\beta}\) is
estimated by solving the estimating equation 
$
\frac{\partial\bm{\mu}}{\partial\bm{\beta}}
\bm{\Omega}
\left( \theta^{\Omega} \right)^{-1}
\left( \bm{y} - \bm{\mu}\left( \bm{\beta} \right) \right) = 0
$.
GEEs yield consistent estimates of \(\bm{\beta}\) and
its sampling variance even if the covariance structure is misspecified
and is non-sparse \citep{xie_asymptotics_2003}.

GEEs can naturally be adapted to case-control settings by using the ascertained conditional mean function
$E\left[y_i | \bm{X}_i,s_i=1,\bm{\beta}\right] = P\left(y_i=1\,|\,\bm{X}_i,s_i=1,\bm{\beta}\right)$. We now show how the GEE fixed effect estimates can be plugged into GPs. In the general case it is not possible to reconcile fixed effect estimates of GEEs and GPs, because GEEs assume that the conditional mean of the outcome is affected only by the fixed effects, whereas GPs assume that it is affected by both the fixed effects and  the GP latent variable. Fortunately, the probit likelihood  provides a convenient way to reconcile the two approaches. 
Denote $\bm{\beta}_{\textrm{GEE}}$ and $\bm{\beta}_{\textrm{GP}}$  as the vectors of fixed effects used by GEE and GP, respectively. When using a probit likelihood, the GEE conditional mean is given by 
$
\Phi(\bm{X}_i^T \bm{\beta}_{\textrm{GEE}})
$,
where $\Phi(\cdot)$
is the standard normal cumulative density. In contrast, the GP conditional mean is given by
$
\Phi\left(\frac{\bm{X}_i^T \bm{\beta}_{\textrm{GP}}}
{\left(\textrm{var}(g_i)+1\right)^{1/2}}\right)
$.
If $\text{var}(g_i)$ is constant for every unit $i$ (which  corresponds to a constant value on the diagonal of the covariance matrix of $\bm{g}$), the two approaches can be reconciled by defining 
$
\bm{\beta}_{\textrm{GP}} = 
\bm{\beta}_{\textrm{GEE}} \left(\textrm{var}(g_i)+1 \right)^{1/2}.
$
In practice, the diagonal of the covariance matrix of $\bm{g}$ is often exactly or almost exactly constant, which enables exploiting the above relation. Therefore, we can use the GEE estimates in a GP by setting
$\bm{\beta}_{\textrm{GP}} = 
\bm{\beta}_{\textrm{GEE}} \left(\textrm{var}(g_i)+1 \right)^{1/2}$.

Our implementation of AGEE closely followed that of \citep{jiang_retrospective_2016}, with a suitable modification of the conditional mean to encode ascertainment, as described above.

\bibliography{zotero_papers2}

\end{document}